\documentstyle[multicol,aps,pre,eqsecnum]{revtex}
\begin{document}
\def\B.#1{{\bbox{#1}}}                  %
\title{
{\rm PHYSICAL REVIEW E\hfill Submitted}\\~~\\
  Towards a Nonperturbative Theory of Hydrodynamic Turbulence:\\
  Fusion Rules, Exact Bridge Relations and Anomalous
 Viscous Scaling Functions}
  \author {Victor L'vov\footnote{Electronic address: 
fnLvov@wis.weizmann.ac.il ~~~~~~~~~~~~~~
URL:~~http://lvov.weizmann.ac.il}  
and Itamar  Procaccia
\footnote{Electronic address: cfProcac@weizmann.weizmann.ac.il,
~~~URL: http://chemphys.weizmann.ac.il/procaccia.html }   }
  \address{$^{*\dagger}$Department of~~Chemical Physics, 
 The Weizmann Institute of Science,
  Rehovot 76100, Israel,\\
  $^*$Institute of Automation and Electrometry,
   Ac. Sci.\  of Russia, 630090, Novosibirsk, Russia}
 \maketitle
 \begin{abstract}
   of equations is decoupled in the sense that the correlations of
   In this paper we address nonperturbative aspects of the analytic
   theory of hydrodynamic turbulence.  Of paramount importance for
   this theory are the ``fusion rules" that describe the asymptotic
   properties of $n$-point correlation functions when some of the
   coordinates tend toward one other. We first derive here, on the
   basis of the Navier-Stokes equations, a set of fusion rules for
   correlations of velocity differences when all the separation are in
   the inertial interval. Using this set of fusion rules we consider
   the standard hierarchy of equations relating the $n$-th order
   correlations (originating from the viscous term in the
   Navier-Stokes equations) to $n+1$'th order (originating from the
   nonlinear term) and demonstrate that for fully unfused correlations
   the viscous term is negligible. Consequently the hierarchic chain
   of equations is decoupled in the sense that the correlations of 
   $n+1$'th order satisfy a homogeneous equation that may exhibit
   anomalous scaling solutions. Using the same hierarchy of equations
   when some separations go to zero we derive a second set of fusion
   rules for correlations with differences in the viscous range. The
   latter includes gradient fields. We demonstrate that every n'th
   order correlation function of velocity differences $\bbox{\cal
     F}_n(\B.R_1,\B.R_2,\dots)$ exhibits its own cross-over length
   $\eta_{n}$ to dissipative behavior as a function of, say, $R_1$.
   This length depends on $n$ {and on the remaining separations}
   $R_2,R_3,\dots$.  When all these separations are of the same order
   $R$ this length scales like $\eta_n(R)\sim \eta (R/L)^{x_n}$ with
   $x_n=(\zeta_n-\zeta_{n+1}+\zeta_3-\zeta_2)/(2-\zeta_2)$, with
   $\zeta_n$ being the scaling exponent of the $n$'th order structure
   function. We derive a class of exact scaling relations bridging the
   exponents of correlations of gradient fields to the exponents
   $\zeta_n$ of the $n$'th order structure functions. One of these
   relations is the well known ``bridge relation" for the scaling
   exponent of dissipation fluctuations $\mu=2-\zeta_6$.
 \end{abstract}
\begin{multicols}{2}
\section{introduction}
In a recent series of papers \cite{LP-0,LP-1,LP-2,LP-3} we developed
an approach to the theory of the universal scaling properties that
characterize the statistical invariants of fully developed turbulent
flows. The main technical tool of this approach was renormalized
perturbation theory which allows, after exact resummations, to obtain
insights about the non-perturbative properties of the universal
statistics of turbulence.  The main nonperturbative results which were
obtained are (i) Hydrodynamic interactions are local in physical-space
and in scale-space once the sweeping effects are removed.  (ii) The
inner, viscous scale which appears perturbatively as an Ultra-Violet
cutoff leads non-perturbatively to the anomalous scaling of
correlations involving velocity gradients. (iii) The outer scale of
turbulence appears in the theory as a renormalization scale after
infinite resummations of a renormalized perturbation series that
converges order by order.  The aim of this paper is to present new
nonperturbative results that together with the previous findings begin
to assemble to a solid structure of the theory of hydrodynamic
turbulence. Brief presentations of some of the new results were
offered in \cite{96LP-1,96LP-2,96LPP,96LP-3}.
 
The issue under study is not new; decades of experimental and
theoretical attention (see for example
\cite{MY-2,84AGHA,93SK,93Ben,Fri}) have been devoted to two types of
simultaneous correlation functions; the first type includes the
structure functions of velocity differences,
\begin{eqnarray}
S_n(\B.R) &=& \left\langle\vert\ {\B.w}({\B.r},{\B.r}')
\vert^n\right\rangle \ , \\
{\B.w}({\B.r},{\B.r}')
&\equiv& {\B.u} ({\B.r}')- {\B.u}({\B.r})\ , \quad \B.R\equiv 
\B.r'-\B.r\label{Sns}
\ ,\label{du1}
\end{eqnarray}
where $\left<\dots\right>$ stands for a suitably defined ensemble
average.  A second type of correlations include gradients of the
velocity field.  An important example is the rate $\epsilon({\bf
  r},t)$ at which energy is dissipated into heat due to viscous
damping. This rate is roughly $ \nu|\nabla {\B.u}({\B.r},t)|^2$.  An
often-studied simultaneous correlation function of 
$\hat\epsilon({\B.r},t) =
\epsilon ({\B.r},t) - \bar\epsilon$ is
\begin{equation}
K_{\epsilon\epsilon}(\B.R) = \left< \hat\epsilon ({\B.r}+{\B.R})
\hat\epsilon ({\B.r}) \right> \ . \label{Kee}
\end{equation}
Within the famous Kolmogorov 1941 (K41) approach \cite{41Kol} to
turbulence one predicts that for $R$ in the inertial range, i.e $\eta
\ll R \ll L$, $S_n(R)$ is $\left (\bar \epsilon R \right )^{n/3}$ and
$K_{\epsilon\epsilon}(R) \simeq \nu^2 \bar \epsilon^{4/3}R^{-8/3}$.
Here $L$ and $\eta$ are respectively the outer scale of turbulence and
the Kolmogorov viscous cutoff. $\bar\epsilon$ is the mean energy flux
per unit time per unit mass.

Experimental measurements show that in some aspects Kolmogorov was
remarkably close to the truth.  The major aspect of his predictions,
that the statistical quantities depend on the length scale $R$ as
power laws, is corroborated by experiments. On the other hand, the
predicted exponents seem not to be exactly realized.  For example, the
correlation $K_{\epsilon\epsilon}(R)$ decays according to a power law,
\begin{equation}
K_{\epsilon\epsilon}(R) \propto R^{-\mu} \ , \label{defmu}
\end{equation}
for $   \eta \ll R \ll L$, with $\mu$ having a numerical value of
$0.2-0.3$ instead of 8/3 \cite{93SK}.  The structure functions also
behave as power laws,
\begin{equation}
S_n(R) \propto R^{\zeta_n} \ , \label{defzn}
\end{equation}
but the numerical values of $\zeta_n$ deviate progressively from $n/3$
when $n$ increases \cite{84AGHA,93Ben}.  Something fundamental seems
to be missing.

The first positive answer within our approach appeared in the context
of correlation functions of gradient fields rather than of the
velocity differences \cite{LP-2,95LL}.  As indicated above the
discrepancy in the exponent $\mu$ between the experimental value and
the naive expectation is large. This gave hope that the explanation
must lie close to the surface. Indeed, considering the perturbative
scheme for $K_{\epsilon\epsilon}(R)$ leads immediately to the
discovery of logarithmic ultraviolet divergences with some ultraviolet
viscous scale $\tilde \eta$ acting as the renormalization scale.  The
summation of this infinite series results in a factor $(R/\tilde
\eta)^{2\Delta}$ with some anomalous exponent $\Delta$ which is,
generally speaking, of the order of unity.  Additional calculations
lead to the exact result $\Delta = 2-\zeta_2$ and to the understanding
that this result means that the renormalized perturbation series for
the structure functions diverges in the limit $L \to \infty$ like
$(L/R)^{\delta_n}$. The anomalous exponents $\delta_n$ are the
deviations of the exponents of $S_n(R)$ from their K41 value.  This is
a very delicate and important point.  It is not simple to see why a
series each of whose terms converges when $L\to \infty$ still diverges
as a whole.  One can understand this as in re-summing the series of a
desired function one finds an inhomogeneous equation whose
inhomogeneous solutions are indeed $L$-independent. However, the
equation possesses also homogeneous solutions which are inherently
nonperturbative in nature and may have anomalous scaling.  Homogeneous
solutions must be matched with the boundary conditions, and this is
the way that the outer scale appears in the theory. An important
remaining step in the theoretical development is to understand how to
compute the anomalous exponents $\delta_n$.

The divergence of the entire perturbation series for $S_n(R)$ with
$L\to \infty$ forces us to seek a nonperturbative handle on the
theory. One fact that may potentially lead to a nonperturbative
control is the existence of a global balance between energy input and
dissipation. This may be turned into a nonperturbative constraint on
each $n$-th order structure function \cite{LP-3}. Using the
Navier-Stokes equations one derives the set of equations of motion
\begin{equation}
{\partial S_n(R,t) \over \partial t}+D_n(R,t) = J_n(R,t) \ ,
\label{dSn}
\end{equation}
where $D_n$ and $J_n$ stem from the nonlinear and the viscous terms in
the Navier Stokes equations respectively. In the stationary state the
time derivative vanishes and one has the balance equation
\begin{equation}
D_n(R) =  J_n(R) \ . \label{baleq}
\end{equation}
The evaluation of $D_n(R)$ does not pose conceptual difficulties. It
was shown \cite{LP-3} that it is of the order of $dS_{n+1}/dR$. The
evaluation of $J_n(R)$ poses some delicate issues, but it was shown
that it is proportional to $S_n(R)/S_2(R)$.  This result is of course
the famous hierarchic chain of equations of which many closure
attempts have failed to provide the desired anomalous scaling
solutions.

At the heart of this paper we offer a way out of this conundrum. The 
problem with Eq.(\ref{baleq}) is that it refers to a ``fully fused"
quantity, and this fact leads to difficulties. What is meant by this
is that the primitive object of the statistical theory of turbulence
is {\em not} the structure function $S_n(R)$, but rather the ``fully
unfused" $n$-rank tensor correlation function of velocity differences:
\begin{eqnarray}
&&{\B.F}_n({\B.r}_1,{\B.r}'_1;{\B.r}_2,{\B.r}'_2;
\dots;{\B.r}_n,{\B.r}'_n)
\nonumber \\
&=& \left< \B.w({\B.r}_1,{\B.r}'_1)
\B.w({\B.r}_2,{\B.r}'_2) \dots
\B.w({\B.r}_n,{\B.r}'_n) \right> \ .
\label{defF}
\end{eqnarray}
In this quantity all the coordinates are distinct. The structure
function $S_n(R)$ is obtained by fusing all the coordinates $\B.r_i$
into one point $\B.r$, and all the coordinates $\B.r'_i$ into another
point $\B.r+\B.R$. In this process of fusion one crosses the
dissipative scale. To control this process we need to formulate the
so-called ``fusion rules" for turbulence, which govern the properties
of $n$-point correlation functions when two or more coordinates
``fuse" together.  One cannot fuse blindly.

The main fundamental results of this paper are as follows:
\begin{enumerate}
\item {\em Fusion rules in the inertial interval}.  The explanation of
  how fusion works was briefly given in \cite {96LP-1}.  We begin this
  paper in Sections II and III with a detailed derivation of the
  fusion rules, considering all the possible relative geometries of
  the sets of fused and unfused points.
\item {\em The homogeneous equation for the $(n+1)$-order correlation
    functions}: the fusion rules allow us to consider in Section IV
  and V the generalized balance equation (\ref{bal}) for the fully
  unfused correlation functions.  We demonstrate that for these
  quantities the hierarchic chain of equations decouples, since now
  the dissipative term vanishes when the viscosity $\nu$ goes to zero.
  In the case of Eq.(\ref{baleq}) the term $J_n$ contains a
  correlation function with many fused coordinates. As a result (and
  cf. Section VI) it remains finite in the limit $\nu \to 0$. In the
  present case of fully unfused quantities we get in the limit $\nu
  \to 0$ a {\em homogeneous equation} involving only correlations of
  order $n+1$. The full analysis of this equation is not easy, and it
  deserves a considerable amount of attention in the future.
\item {\em The existence of ``viscous scaling functions"}.  Having
  obtained the generalized balance equation we can now study precisely
  when, as a function of one separation distance (say
  $|\B.r'_1-\B.r_1|$), the fully unfused correlation function crosses
  over to dissipative behavior. It turns out that the scale at which
  this happens depends on the rank $n$ {\em and on the remaining
    inertial range distances}. The dependence on the inertial range
  distances is characterized by anomalous exponents, and we compute
  them exactly in terms of the anomalous exponents $\zeta_n$. This is
  done in Section VI.
\item {\em Exact bridge relations}. With a precise evaluation of the
  viscous scaling functions and the fusion rules in hand we can
  consider correlation functions of gradient fields, and relate their
  scaling exponents to the anomalous exponents $\zeta_n$. The result
  of this exercise, which is presented in Section VII, is an infinite
  set of bridge relations. We consider correlations of the type \FL
\begin{eqnarray}
\B.{\cal K}_{p\epsilon}^{(n)}&\equiv& \left<\epsilon(\B.x_1)
\epsilon(\B.x_2)\dots
\epsilon(\B.x_p)
\B.w(\B.r_1,\B.r'_1)\!\dots \!\B.w(\B.r_n,\B.r'_n)\right>\nonumber\\
&&\propto R^{-\mu^{(p)}_n}, 
\end{eqnarray}
where $R$ is a typical separation between any pair, and we are
interested in the scaling relations between the exponents
$\mu^{(p)}_n$ and the exponents $\zeta_n$. Note that $\mu^{(2)}_0$ in
this notation is the well studied \cite{92Pra,93SK} exponent of
dissipation fluctuation which is denoted $\mu$. On the basis of the
Navier-Stokes equations and the fusion rules we establish rigorously
that
\begin{equation}
\mu^{(p)}_n = p-\zeta_{n+3p} \ . 
\end{equation}
In particular we offer a solid derivation of the phenomenologically
guessed bridge relation \cite{MY-2,Fri} $\mu=2-\zeta_6$.
\end{enumerate}
In Section VIII we summarize the paper and indicate the direction of
the road ahead.
\section{Fusion rules for the inertial interval}
\label{FRII}
In this section we discuss the fusion rules that arise when some of
the coordinates of an $n$-point correlation function tend toward one
other. We begin by introducing the set of correlation functions that
are required by the analysis.
\subsection{Correlation functions and related quantities}
The fully unfused correlation function ${\B.F}_n$ was introduced in
Eq.(\ref{defF}).  Other statistical quantities of interest have the
same number of velocity differences but they depend on a smaller
number of coordinates. In the language developed below we will refer
to such quantities as partially fused correlation functions.  For
example if all the velocity differences are evaluated with respect to
the same point we define an $n$-rank tensor \FL
\begin{equation}
{\B.S}_n({\B.r}_0|{\B.r}_1,{\B.r}_2\!\dots\!{\B.r}_n) \!=\!
\left<\B.w({\B.r}_0,{\B.r}_1)
\B.w({\B.r}_0,{\B.r}_2)\dots \B.w({\B.r}_0,{\bf
r}_n)\right> .
\label{Sn}
\end{equation}
Note that in an inhomogeneous ensemble this quantity depends
explicitly on $\B.r_0$. Obviously, ${\B.F}_n$ depends on many more
coordinates than ${\B.S}_n$. It is useful sometimes, therefore, to
represent correlation functions ${\B.F}_n$ in terms of the partly
fused quantities ${\B.S}_n$. This can always be done: represent
$\B.w({\B.r}_j,{\B.r}'_j)$ as $\B.w({\B.r}_j,{\B.r}_0)+
\B.w({\B.r}_0,{\B.r}'_j)$, and by substitution find the needed
relation.  It can also be useful to choose $\B.r_0$ as the center of
mass of all the coordinates in ${\B.F}_n$. We will return to such
relations when we need them.

One can fuse more points together. For example we can have only one
separation vector distance, and form the $n$-rank tensor
\begin{equation}
\hat \B.S_n(\B.R)\equiv \hat{\B.S}_n({\B.r}_0|{\B.r}) =
 {\B.S}_n({\B.r}_0|{\B.r},{\B.r}\dots{\B.r}) \ ,
\label{SnR}
\end{equation}
where $\B.R\equiv \B.r-\B.r_0$.  Equations (\ref{Sn}) and (\ref{SnR})
can be understood as the limit of (\ref{defF}) when pairs of
coordinates fuse together. For example, in going from (\ref{defF}) to
(\ref{Sn}) we fuse all the unprimed coordinates, and keep all the
primed coordinates different. To get (\ref{SnR}) we also fuse all the
primed coordinates.

Note that one can also consider another type of fusion, in which pairs
of coordinates $\B.r_j$ and $\B.r'_j$ which are associated with the
velocity difference $\B.w(\B.r_j,\B.r'_j)$ coalesce. We will refer to
the fusion of coordinates with a velocity difference as a fusion of
type A, and a fusion of coordinates that are not associated with a
velocity difference as a fusion of type B.  The properties of these
two limiting procedures are different, and they have to be studied
separately. It is commonly assumed that the limit of type B exists,
and the limit remains finite when $\nu\to 0$.  This is equivalent to
the statement that ${\B.S}_n({\B.r}_0|{\B.r}_1,{\bf
  r}_2\dots{\B.r}_n)$ and ${\B.S}_n({\B.R})$ are independent of Re for
asymptotically large Re. Although this property looks innocent, it is
a deep statement that in general needs to be supported by theory. In
the language of diagrammatic perturbation theory this property is a
consequence of the locality of every diagram in the perturbation
expansion for these quantities; this property was discussed in detail
in \cite{LP-1}. In \cite{LP-2} it was explained that there exists a
mechanism for the appearance of the viscous scale in the statistical
theory.  However, this mechanism operates only when one correlates
gradients of the velocity fields rather than the velocity differences
themselves.  Thus the above statement is equivalent to the {\em
  assumption}\/ that there are no additional non-perturbative
mechanisms for the appearance of the viscous scales in the theory of
correlations of velocity differences.  Additional objects that appear
naturally in the theory are correlations whose tensorial nature is of
lower rank, including scalars and vectors. The first one is the scalar
quantity which is appropriate for even orders of ${\B.S}_n$. To keep
in mind its scalar nature we denoted it in \cite{LP-3} as
$\stackrel{o}{S}_{2m}(R)$ and define it as
 \begin{equation}
 \stackrel{o}{S}_{2m}(\B.R) \equiv \left< |\B.w({\B.r}_0,{\B.r})
 |^{2m}\right>
 \,,\quad {\B.R}\equiv {\B.r}-{\B.r}_0 \ .
 \label{defSsc}
 \end{equation}
 The quantity $\stackrel{o}{S}_{2m}(R)$ is analytic. For odd order
 structure functions we introduce a vector object
 $S^{\alpha}_{2m+1}({\B.R})$ according to
 \begin{equation}
 S^{\alpha}_{2m+1}({\B.R})\equiv
\left<w_{\alpha}({\B.r}_0,{\B.r})
 |\B.w({\B.r}_0,{\B.r}) |^{2m}\right> \ .
\label{defSvec}
 \end{equation}
 Here and below we will use Greek indices to indicate vector and
 tensor components, and Roman indices to indicate the order of the
 quantity. The placement of indices as subscripts or superscripts has
 no special meaning, and is chosen for convenience.
 
 In isotropic turbulence the vector $S^{\alpha}_{2m+1}({\B.R})$ can
 only be oriented along ${\B.R}$. This allows us the introduction of a
 scalar quantity $S_{2m+1}(R)$ which depends on the magnitude of $R$:
\begin{equation}
 S^{\alpha}_{2m+1}({\B.R}) = {R_\alpha \over R}S_{2m+1}(R) \ . 
\label{S2m1}
 \end{equation}
 In later sections we use the objects defined here to study the fusion
 rules and their consequences.
\subsection{Basic Properties of Correlation Functions in Systems with
  Flux Equilibrium} 

The systems that we discuss are driven on a characteristic scale
referred to as the outer scale $L$. This driving can be achieved by
either a time dependent low frequency ``stirring force" or by
specifying given values of ${\B.u}$ at a set of ``boundary" points
with a characteristic separation $L$ away from our observation points
${\B.r}_0,{\B.r}'_0, {\B.r}_1$, etc. The system suffers dissipation
due to viscosity, and in the inviscid limit the kinetic energy is
conserved. In three dimensions we deal with a ``direct" energy cascade
in which the intake of energy on the scale $L$ is balanced by
dissipation on a small scale $\eta\ll L$.

We invoke two fundamental assumptions \cite{41Kol}:

\noindent
{\B.1.} {\em Scale invariance}\/: all the correlation functions are
homogeneous functions of their arguments in the core of the inertial
interval $\eta \ll |{\B.r}_i -{\B.r}_0| \ll L$:
\begin{equation}
  {\B.F}_n(\lambda{\B.r}_1,\lambda{\B.r}'_1 ;\dots ;\lambda{\bf
    r}_n,\lambda{\B.r}'_n) = \lambda^{\zeta_n}{\B.F}_n({\bf
    r}_1,{\B.r}'_1;\dots;{\B.r}_n,{\B.r}'_n) \ , \label{assum2}
\end{equation}
where $\zeta_n$ are scaling exponents. It is obvious that all the
other types of correlation and structure functions that emanate from
this most general quantity and which were detailed in the previous
subsection share the same scaling exponents.  Accordingly $\zeta_n$ is
also the usual scaling exponent of the $n$-th order structure function
(\ref{defSsc}).

\noindent
{\B.2.} {\em Universality of the scaling exponents}\/: this means that
we can fix an arbitrary set of velocity differences on the scale of
$L$, and the scaling exponents of the correlation functions will not
depend on the precise choice of the $L$-scale motions.  Mathematically
this is expressed as the following property of the conditional
average: if for $i \leq n$ $|{\B.r}_i -{\B.r}'_i| \ll L$, and for $i >
n$, $|{\B.r}'_i -{\B.r}_i| \sim L$, then 
\FL
\begin{eqnarray}
&&\big\langle \B.w({\B.r}_1,{\B.r}'_1)\B.w({\bf
r}_2,{\B.r}'_2) \dots
\B.w({\B.r}_n,{\B.r}'_n) \big |\B.w({\B.r}_{n+1},{\bf
r}'_{n+1}) \nonumber \\
&&\dots \B.w({\B.r}_{n+N},{\B.r}'_{n+N}) \big\rangle =
{\tilde{\B.F}}_n({\B.r}_1,{\B.r}'_1;
\dots;{\B.r}_n,{\B.r}'_n) \nonumber \\
&\times& \Phi_{n,N} ({\B.r}_{n+1},{\B.r}'_{n+1};\dots ; {\B.r}_{n+N},
{\B.r}'_{n+N})
\label{assum1}
\end{eqnarray}
The precise meaning of the universality
assumption is that the functions ${\tilde{\B.F}}_n$ have the same
scaling exponents as ${\B.F}_n$ in the inertial interval. They may be
different functions in the inertial interval and in particular they
may differ in their crossover to viscous behavior.  Their (different)
crossover scales may depend on the large scale motions that were fixed
in the conditional average.

\subsection{Fusion rules for the fusion of $p$ pairs of points in
velocity differences (type A)}
\subsubsection{General case: $2 \le p \le n-2$}
The first set of fusion rules that we derive concerns ${\B.F}_n$ when
$p$ pairs of points $\B.r_1,\B.r'_1\dots \B.r_p,\B.r'_p$, $(p<n)$
which involve $p$ velocity differences tend to some point ${\B.r}_0 $,
see Fig.\ref{Fig1}. Here we exclude the special cases $p=1$ and
$p=n-1$ in which the leading scaling contribution vanishes by
symmetry. We will consider these special cases in the next
subsections.

Consider the situation in which the typical separation of all the
coordinates $\B.r_1,\dots,\B.r'_p$ from ${\B.r}_0$ is $r$, whereas all
the other separations remain much larger, say of the order of $R$,
$r\ll R \ll L$. We will show, on the basis of assumptions $\B.1$ and
$\B.2$, that
\begin{eqnarray}\label{fusion1}
&& {\B.F}_n({\B.r}_1,{\B.r}'_1;
\dots;{\B.r}_n,{\B.r}'_n)\\
&=&
{\tilde{\B.F}}_p({\B.r}_1,{\B.r}'_1;
\dots;{\B.r}_p,{\B.r}'_p)
\bbox\Psi_{n,p}
({\B.r}_{p+1},{\B.r}'_{p+1};\dots;{\B.r}_n,{\B.r}'_n) \ ,
 \nonumber 
\end{eqnarray}
where ${\tilde{\B.F}}_p$ is a tensor of rank $p$ associated with the
first $p$ tensor indices of ${\B.F}_n$. The $(n-p)$-rank tensor
$\bbox\Psi_{n,p}({\B.r}_{p+1},{\B.r}'_{p+1};\dots;{\B.r}_n,{\bf
  r}'_n)$ is a homogeneous function with a scaling exponent
$\zeta_n-\zeta_p$, and is associated with the remaining $n-p$ indices
of ${\B.F}_n$. The derivation of the fusion rule (\ref{fusion1})
follows from Bayes' theorem. We write
\FL
\begin{eqnarray}
&{\B.F}&_n({\B.r}_1,{\B.r}'_1;\dots{\B.r}_n,{\B.r}'_n)=\!\int \!
d\B.w({\B.r}_{p+1},{\B.r}'_{p+1})\!\dots  d\B.w({\bf
r}_n,{\B.r}'_n)\nonumber\\
&\B.w&({\B.r}_{p+1},\!{\B.r}'_{p+1})\!\dots  \B.w({\bf
r}_n,\!{\B.r}'_n)
{\cal P}[\B.w({\B.r}_{p+1},\!{\B.r}'_{p+1})\! 
\dots \B.w({\B.r}_n,\!{\B.r}'_n) ]
\nonumber \\
&\times&\big\langle \B.w({\B.r}_1,{\B.r}'_1),
\B.w({\B. r}_2,{\B.r}'_2)\dots
\B.w ({\B.r}_p,{\B.r}'_p)\Big |\B.w
({\B.r}_{p+1},{\B.r}'_{p+1})\nonumber \\
&\times& \B.w({\B.r}_{p+2},{\B.r}'_{p+2})\dots
\B.w({\B.r}_n,{\B.r}'_n) \big\rangle \ ,
\label{bayes}
\end{eqnarray}
\noindent
where ${\cal P}[\B.w({\B.r}_{p+1},{\B.r}'_{p+1})\dots
\B.w({\B.r}_n,{\B.r}'_n) ]$ is the probability to see the tensor
$\B.w({\B.r}_{p+1},{\B.r}'_{p+1})\dots \B.w({\B.r}_n,{\B.r}'_n)$.
Next note the consequence of assumption {\B.2}: the scaling laws of
the correlation functions at scale $r$ are the same independent of
whether we force the system on the scale $L\gg r$ or on the scale
$R\gg r$. The conditional average in (\ref{bayes}) is proportional to
${\tilde{\B.F}}_p$, and hence (\ref{fusion1}).

\begin{figure}
\label{Fig1}
\end{figure}

{\small
FIG.~\ref{Fig1}
The geometry of fusion of type A. Lines connecting points indicate
  velocity differences across that distance. In this example there are
  three velocity differences across small separations (of the order of
  $r$) and two velocity differences across large separations, of the
  order of $R$.}\vskip .4cm

The first impression is that (\ref{fusion1}) means statistical
independence of the small scale motion from the large scales. This is
not so. The difference is that the assumption of statistical
independence would lead to $\B.F_n=\B.F_p \B.F_{n-p}$. Indeed, the
first factor $\B.F_p$ has the same order of magnitude and the same
exponent as $\tilde \B.F_p$ in (\ref {fusion1}).  However, the factor
$\B.F_{n-p}$ has a scaling exponent $\zeta_{n-p}$ rather than
$\zeta_n-\zeta_p$ which is the exponent of $\B.\Psi_{n,p}$. When all
the large separation are of the same order $R$, $\B.\Psi_{n,p}\sim
S_n(R)/S_p(R)$ which in the case of multi-scaling is much larger than
$\B.F_{n-p}\sim S_{n-p}(R)$. We thus understand that the fusion rules
in fact demonstrate the existence of a very special statistical
dependence of the small scales on the large scales. This dependence
stems physically from the existence of a direct energy flux from large
to small scales. We will see that it can lead to a totally
unconventional scaling structure of the theory. We should stress that
these fusion rules were derived from first principles for $p=2$ in
Navier-Stokes turbulence \cite{LP-3} and for passive scalar advection
for any $p$ \cite{96FGLP}.
\subsubsection{Fusion of one pair of points in a velocity difference}
\label{fusep1}
As we mentioned the case $p=1$ in which the velocity difference across
a small scale $r\equiv|\B.r_1-\B.r'_1|$ is correlated with $n-1$
velocity differences across larger distances requires a special
attention. A naive application of the fusion rule (\ref{fusion1})
results in $\B.F_n \propto \B.F_1({\B.r}_1,{\B.r}'_1)$ which
vanishes due to space homogeneity. In order to evaluate the
leading non-vanishing term we expand $\B.u(\B.r'_1)$ in a Taylor
series around $\B.u(\B.r_1)$,
\begin{equation}
\B.u(\B.r'_1)=\B.u(\B.r_1)
+\bbox\nabla_1\B.u(\B.r_1)\cdot(\B.r'_1-\B.r_1)+\dots \ . \
\end{equation}
Using this we can write
\begin{eqnarray}
&&\lim_{\B.r_1\to\B.r'_1} {\B.F}_n({\B.r}_1,{\B.r}'_1;\dots;{\bf
r}_n,{\B.r}'_n) \nonumber \\
&&=(\B.r'_1-\B.r_1)\cdot\bbox\nabla_1\left<\B.u(\B.r_1)\B.w({\bf
r}_2,{\B.r}'_2)
\dots\B.w({\B.r}_n,{\B.r}'_n)\right> \ .  \nonumber
\end{eqnarray}
The correlation function in this formula depends an all the separation
distances, and the gradient with respect to $\B.r_1$ picks up
contributions from all the differences $\B.r_j-\B.r_1$. Therefore the
gradient can be evaluated as the inverse of the smallest of these,
$|\bbox\nabla_i|\sim 1/R_{\rm min}$ where $R_{\rm min}\equiv
\min_i\{|\B.r_i-\B.r_1|\}$. For $r\ll R_{\rm min}$ this leads to the
evaluation
\begin{equation}
{\B.F}_n({\B.r}_1,{\B.r}'_1;\dots;{\B.r}_n,{\B.r}'_n) \sim
{r\over R_{\rm min}}S_n(R) \ . \
\label{rule2}
\end{equation}
This formula creates an immediate worry about the situation in which
$R_{\rm min}$ becomes of the order of $r$ or smaller. We will analyze
this situation in Subsection ~\ref{sub-D}. Now we will consider the
next special case of fusion, when $p=n-1$.
\subsubsection{Fusion rule for one large separation distance}
Consider ${\B.F}_n({\B.r}_1,{\B.r}'_1;{\B.r}_2,\B.r'_2;\dots;
{\B.r}_n,{\B.r}'_n)$ with all the coordinates being nearby except for
${\B.r}_1$ which is far away, a distance $R$ from the remaining $2n-1$
coordinates which are all within a ball of radius $r$.  Since we
assume that the flow is isotropic, the tensor ${\B.F}_n$ is a
(generally reducible) representation of the rotation group. Moreover,
in the situation discussed here the isotropization of the small scales
(of the order of $r$) with respect to the direction of $\B.R\equiv
{\B.r}'_1-{\B.r}_1$ leads to a direct product structure
\begin{eqnarray}
 &&F^{\alpha_1\alpha_2\dots\alpha_n}_n({\B.r}_1,{\B.r}'_1;{\B.r}_2,{\bf
r}'_2;
\dots;{\B.r}_n,{\B.r}'_n)\nonumber \\&=&
C  {R^{\alpha_1}\over R} \tilde F^{\alpha_2\dots\alpha_n}_n(R,{\B.r}_2,
{\B.r}'_2;\dots;{\B.r}_n,{\B.r}'_n) \ . \label{prst}
\end{eqnarray}
The constraint of incompressibility, which is always true, can be
written as
\begin{equation}
\nabla^{\alpha_1}_1 F^{\alpha_1\alpha_2\dots\alpha_n}_n({\B.r}_1,{\bf
r}'_1;{\B.r}_2,{\B.r}'_2;
\dots;{\B.r}_n,{\B.r}'_n)=0 \ . \
\end{equation}
Applying the divergence operator to (\ref{prst}) we find the following
form of the incompressibility constraint.
\begin{equation}
C\left(2+R{\partial \over\partial R}\right)\tilde
F^{\alpha_2\dots\alpha_n}_n(R,{\B.r}_2,
{\B.r}'_2;\dots;{\B.r}_n,{\B.r}'_n)=0 \ . \
\end{equation}
This equation has two solutions:(i) $C=0$, (ii) $\tilde F_n\propto
 1/R^2$.  However according to the general rule $\tilde F_n$ must be
proportional to $R^{\zeta_n-\zeta_{n-1}}$. Thus the first solution is
realized, the contribution which is $\propto R^{\zeta_n-\zeta_{n-1}}$
vanishes due to the incompressibility constraint. Therefore we need to
consider the next non-vanishing evaluation which is $R$-independent:
\begin{equation}
F^{\alpha_1\alpha_2\dots\alpha_n}_n({\B.r}_1,{\B.r}'_1;{\B.r}_2,{\bf
r}'_2; \dots;{\B.r}_n,{\B.r}'_n)\propto r^{\zeta_n}\ .\label{fuse33}
\end{equation}
This result is interesting. It means that in the case considered here
in which all the separations of order $r$ except one (which is of
order $R$, $\eta\ll r\ll R\ll L$) the scaling exponent of the
correlation function $F_n$ is fully determined by small $r$-scale
fluctuations.

In addition to these cases one needs some special geometries of
fusions of type A in which the general evaluation is inapplicable.
These cases are treated in Appendix A and referred to as needed.
\vskip .2cm

\subsection{Rules for Fusions of Type B}
\label{typeB}
\begin{figure}
\label{Fig2}
\end{figure}

{\small FIG.~\ref{Fig2} The geometry of fusion of type B. Lines
  connecting points indicate velocity differences across that
  distance. In this example there are three velocity differences
  across large separations, but three coordinates within a ball of
  small radius $r$.}
\subsubsection{Coalescence of two points ($p=2$)}
\label{typeB2}
Fusions of type B refer to situations in which there are $p$
coordinates within a ball of small radius $r$, but no two coordinates
that belong to a velocity difference, see Fig~\ref{Fig2}. The simplest
case is $p=2$ in which $|\B.r_1-\B.r_2|$ is much smaller than any
other separation.  Consider the velocity differences
$\B.w(\B.r_1,\B.r'_1)$ and $\B.w(\B.r_2,\B.r'_2)$.  They can be
reexpressed as
\begin{eqnarray}
\B.w(\B.r_1,\B.r'_1)&=&\B.w(\B.r_1,\B.r_0)+\B.w(\B.r_0,\B.r'_1)\ ,
\nonumber\\
\B.w(\B.r_2,\B.r'_2)&=&\B.w(\B.r_0,\B.r'_2)+\B.w(\B.r_2,\B.r_0) \ , 
\label{reex}
\end{eqnarray}
where $\B.r_0=[\B.r_1+\B.r_2]/2$. This allows us to write
\begin{eqnarray}
&&\!\!\langle\B.w(\B.r_1,\!\B.r'_1)\B.w(\B.r_2,\!\B.r'_2)
\{w\}^{n-2}\rangle
=\langle\B.w(\B.r'_1,\!\B.r_0)
\B.w(\B.r'_2,\!\B.r_0)\{w\}^{n-2}\rangle\nonumber \\
 &&\!\!-\langle\B.w(\B.r_1,\!\B.r_0)
\B.w(\B.r_2,\!\B.r_0)\{w\}^{n\!-\!2}\rangle
-\left<\B.w(\B.r'_1,\!\B.r_0)\B.w(\B.r_2,
\!\B.r_0)\{w\}^{n\!-\!2}\right> 
\nonumber \\&&\!\!\!+\left<\B.w(\B.r_1,\B.r_0)
\B.w(\B.r_2,\B.r_0)\{w\}^{n-2}\right>\ . \
\end{eqnarray}
Here we used the short-hand notation $\{w\}^{n-2}$ to denote the
remaining product of $n-2$ velocity differences.

The first term on the RHS is independent of the small separation, and
is a homogeneous function of the large separations with a scaling
exponent $\zeta_n$. The next two terms contain one velocity difference
across small separations which according to the discussion of Fig.~
is proportional to $r^{\zeta_2}$. The last term has two velocity
differences across a small separation, and is also proportional to the
same factor. This allows one to formulate a fusion rule of type B for
fusion of two points (not associated with velocity difference) in the
following form:
\begin{eqnarray}
&&\B.F_n({\B.r}_1,{\B.r}'_1;{\B.r}_2,{\B.r}'_2;
\{{\B.r}_k,{\B.r}'_k\})-\B.F_n({\B.r}_1,{\B.r}'_1;{\B.r}_1,{\B.r}'_2;
\{{\B.r}_k,{\B.r}'_k\})
\nonumber\\
&&=\tilde\B.S_2(|\B.r_1-\B.r_2|)
\bbox{\Psi}_{n,2}({\B.r}_0;{\B.r}'_1,{\B.r}'_2;
\{{\B.r}_k,{\B.r}'_k\}) \propto\Big({r\over R}\Big)
^{\zeta_2}R^{\zeta_n} \ .
\nonumber \\
\label{fuse3}
\end{eqnarray}
\subsubsection{Coalescence of three points ($p=3$)}
\label{typeB3}
The next topic of discussion is the fusion of type B (without velocity
differences) of three points, (say $\B.r_1,\B.r_2$, and $\B.r_3$), as
shown in Fig.~\ref{Fig2}. As before the separations between these
points are all of the order of $r$, and all the other separations are
much larger, of the order of $R$, $R\gg r$. As in the case of $p=2$ we
denote the center of mass of these coordinates as $\B.r_0$:
$\B.r_0=(\B.r_1+\B.r_2+\B.r_3)/3$. We express the two of the velocity
differences according to (\ref{reex}), and the third velocity
difference in analogy:
\begin{equation}
\B.w(\B.r_3,\B.r'_3)=\B.w(\B.r_3,\B.r_0)+\B.w(\B.r_0,\B.r'_3) \ . 
\label{reex1}
\end{equation}
Clearly, the correlation function
\begin{eqnarray}
&&\B.F_n({\B.r}_1,{\B.r}'_1;{\B.r}_2,{\B.r}'_2;{\B.r}_3,{\B.r}'_3;
\{{\B.r}_k,{\B.r}'_k\})\nonumber \\
&&=\left<\B.w(\B.r_1,\B.r'_1)
\B.w(\B.r_2,\B.r'_2)\B.w(\B.r_3,\B.r'_3)
\{w\}^{n-3} \right>
\end{eqnarray}
has three types of contributions. The first, $\B.F_n^{(0)}$, is
independent of the small separations,
\begin{equation}
\B.F_n^{(0)}=\B.F(\B.r_0,\B.r'_1;\B.r_0,\B.r'_2;\B.r_0,
\B.r'_3;\{\B.r_k,\B.r'_k\})
\ . \label{Fn0}
\end{equation}
The second type of contribution has either one velocity difference or
a product of two velocity differences across a small separation. Both
these contributions have a leading term that is proportional to
$r^{\zeta_2}$. The third type of contribution has a product of three
velocity differences across a small separation and is proportional to
$r^{\zeta_3}$. In conclusion, the scaling behavior is similar to the
one displayed in Eq.(\ref{hori}), with the addition of a term constant
in $r$, $\B.F_n^{(0)}$.
\subsubsection{General case: fusion of $n$ points without
velocity difference}
\label{typeBn}
The above discussion of fusion of type B of two and three points
allows us to offer a general statement concerning the asymptotic
behavior of $n$-point correlation functions $F_n$ when $p$ of the
coordinates are separated from each other by small distances of the
order of $r$, and $2n-p$ coordinates are separated by a large
separation of the order of $R$. In terms of the definition
(\ref{defF}) we will take the $p$ close-by coordinates as
$\B.r_1,\B.r_2,\dots,\B.r_p$, and the $2n-p$ coordinates as
$\B.r_{p+1},\dots,\B.r_n$ and $\B.r'_1,\dots,\B.r'_n$.  Recall that
this choice means that we do not have velocity differences across
small scales.

To see the asymptotic behavior we need to repeat the substitutions of
the type (\ref{reex}) and (\ref{reex1}), i.e. for $1 \le j \le p$
\begin{equation}
\B.w(\B.r_j,\B.r'_{j'})=\B.w(\B.r_j,\B.r_0)+
\B.w(\B.r_0,\B.r'_{j'}) \ . \label{reex2}
\end{equation}
As before the center of mass is $\B.r_0=(\sum_{k=1}^p \B.r_k)/p$. The
result of this substitution can be readily guessed from the cases
$p=2,3$:
\begin{equation}
\B.F_n({\B.r}_1,{\B.r}'_1;\dots;{\B.r}_p,{\B.r}'_p;
\{{\B.r}_k,{\B.r}'_k\})=\B.F_{n,p}^{(0)}+\sum_{j=2}^p \B.F_{n,p}^{(j)} 
\ . \label{fuseB}
\end{equation}
In this formula $\B.F_{n,p}^{(0)}$ is independent of the small
separation, and each term $\B.F_{n,p}^{(j)}$ originates from $j$
velocity differences across small separations. These velocity
differences are $\B.w(\B.r_i,\B.r_0)$ where $i\le p$. There are
always $C_p^j=p!/j!(p-j)!$ different products of $j$ velocity
differences across small scales in $\B.F_{n,p}^{(j)}$. More
explicitly,
\begin{eqnarray}
\B.F^{(2)}_{n,p}&=&\left[\sum_{i_1>i_2=1}^p S_2
(\B.r_0|\B.r_{i_1},\B.r_{i_2}) \right]\B.\Psi_{n,2}\,,
\label{fuseBp2}\\
\B.F^{(3)}_{n,p}&=&\left[\sum_{i_1>i_2>i_3=1}^p \tilde 
\B.S_3(\B.r_0|\B.r_{i_1},\B.r_{i_2},\B.r_{i_3})\right]
\B.\Psi_{n,3}\ , \label{fuseBp3}
\end{eqnarray}
etc. On the LHS of these equations the functions $\B.F^{(j)}_{n,p}$
depend on all the coordinates. The function $\tilde \B.S_j$ is a
homogeneous function of $j$ separations $\B.r_i-\B.r_0$, with a
scaling exponent $\zeta_j$.  The functions $\B.\Psi_{n,j}$ are
homogeneous functions of $n-p$ large separations $\B.r_k-\B.r_0$ (with
$k=p+1,\dots,n$ and $n$ large separations $\B.r'_m-\B.r_0$ (with
 $m=1,\dots,n)$ with a scaling exponent $\zeta_n-\zeta_j$.
Schematically we can summarize the scaling behavior of
$\B.F^{(j)}_{n,p}$ as
\begin{equation}
\B.F^{(j)}_{n,p}\propto \left({r\over R}\right)^{\zeta_j}
 R^{\zeta_n} \ . 
\label{nicer}
 \end{equation}
 We stress that since we are concerned with the limit $r\ll R$, the
 leading term in (\ref{fuseB}) is always $\B.F^{(0)}_{n,p}$ which is
 independent of $r$.  The leading $r$ dependence is carried by
 $\B.F^{(2)}_{n,p}\propto r^{\zeta_2}$.  This does not mean however
 that the higher order terms in (\ref{fuseB}) are unimportant. They
 will provide the leading order contributions to correlation functions
 of velocity {\em gradients}\/ as will be shown in subsection 2G.

\subsection{Mixture of A- and B-type fusions}
\label{ABfusion}
\begin{figure}
\label{Fig6}
\end{figure}

{\small FIG.~\ref{Fig6} The geometry of fusion with a mixture of type
  A and B}\vskip .4cm

In this Subsection we consider the even more general case in which we
have within the ball of small radius $r$ $q$ pairs of points
associated with the velocity differences $\B.w(\B.r_i,\B.r'_i)$ ($1\le
i \le q$), and $(p-q)$ points (say $\B.r_{q+1},\dots,\B.r_p)$
associated with velocity differences across large separations of order
$R\gg r$, see Fig~\ref{Fig6}. The rest of the coordinates
$\B.r_k,~\B.r'_k,~ k>p$ are separated by large distances of the order
of $R$, see Fig.6. For this case we need again to reexpress $(p-q)$
velocity differences $\B.w(\B.r_j,\B.r'_j)$ (for $j =q+1,\dots,p$) in
the manner of (\ref{reex2}), and to substitute these into the
definition of $\B.F_n$ (\ref{defF}). The result of this process is an
equation similar to (\ref{fuseB}), but with important differences:
\begin{equation}
\B.F_n({\B.r}_1,{\B.r}'_1;\dots;{\B.r}_p,{\B.r}'_p;
\{{\B.r}_k,{\B.r}'_k\})=\sum_{j=q}^p \tilde \B.F_{n,p}^{(j)} 
\ . \label{fuseAB}
\end{equation}
There is no constant term now, and the leading order contribution is
proportional to $r^{\zeta_q}$. The exception (as always) is that when
$q=1$ the leading scaling behavior is $r^{\zeta_2}$. We note that the
functions $\tilde \B.F_{n,p}^{(j)}$ are different from the analogous
functions $\B.F_{n,p}^{(j)}$, but they have the same scaling behavior,
$\tilde \B.F_{n,p}^{(j)}\propto R^{\zeta_n}(r/R)^{\zeta_j}$.  The
explicit expressions analogous to (\ref{fuseBp2},\ref{fuseBp3}) can
be written down when needed. We will find that the subleading terms
contribute the most important contributions in various situations.
\subsection{Fusion rules for the fusion of two or more groups 
  of pairs} The next set of fusion rules is obtained for the structure
function ${\B.F }_n$ when two groups of $p$ and $q$ points (with $p+q
< n$) tend to ${\B.r}_0$ and ${\B.r}'_0$ respectively. The separation
between these groups of points is of the order of $R$. The derivation
of the fusion rules of type A for the simplest situation when all the
coordinates are different (and separated by $r$ or by $R$) obviously
follows from the same basic properties of velocity correlation
functions which we discussed in Subsection 2B. The result looks similar
to Eq. (\ref{fusion1}):
\begin{eqnarray}
\label{fusion2}
&& {\B.F}_n({\B.r}_1,{\B.r}'_1;\dots;{\B.r}_n,{\B.r}'_n)
 \\
&=&{\tilde{\B.F}}_p({\B.r}_1,{\B.r}'_1;\dots;{\B.r}_p,{\B.r}'_p)
{\tilde{\B.F}}_q({\B.r}_{p+1},{\B.r}'_{p+1};\dots;{\B.r}_{p+q},{\bf
r}'_{p+q})
\nonumber \\ &\times&\bbox
\Psi_{n,p,q}({\B.r}_{p+q+1},{\B.r}'_{p+q+1};
\dots;{\B.r}_n,{\B.r}'_n) \ .
\nonumber
 \end{eqnarray}
 The scaling exponent of $\bbox\Psi_{n,p,q}$ is
 $\zeta_{n}-\zeta_p-\zeta_q$.  As in the case of the fusion rules
 (\ref{fusion1}), also (\ref{fusion2}) are {\it not}\/ decompositions
 into products of lower order correlation functions, and the functions
 $\bbox\Psi$ are not correlations of velocity differences across large
 separations. As before the functions $\bbox\Psi$ are much larger
 than the corresponding correlation functions in all situations with
 multi-scaling.  Evidently one can derive similar fusion rules for
 three, four or more groups of coalescing points with large
 separations between the groups. The structure of the resulting
 correlation function will be a product of the correlation function
 associated with each group times some function $\B.\Psi$ of big
 separations which carries the overall exponent.
 
 The generalization of these fusion rules (which are of type A) to the
 more complicated cases with fusions of type B or to mixed types of
 fusion is now obvious: we can consider every group of point
 separately in the way that we discussed for the case of fusion of
 just one group of points.
\subsection{Fusion rules for correlation 
  functions including unfused velocity gradients} In this subsection
we use the fusion rules obtained above to evaluate the leading order
scaling behavior of correlation functions that include unfused
velocity derivatives. To be specific, consider the $q$-order
derivatives $\bbox\nabla_1 \bbox\nabla_2\dots\bbox\nabla_q$, with
~$\bbox\nabla_j\equiv {\partial/ \partial \B.r_j}$. We are going to
apply this $q$-order derivative on correlation functions with $p$
fusing points, with $q\le p$, such that the derivatives operate only
on coordinates within this group of $p$ points.  We will also consider
a constrained derivative, i.e. such that
$\bbox\nabla_1+\bbox\nabla_2+\dots+\bbox\nabla_q=0$. With such a
derivative we get a particularly simple result. There are two
situations to consider. If the $p$ fusing points undergo a fusion of
type A, the leading scaling behavior of the $q$-order derivative is
simply $r^{\zeta_p-q}$. In the case of fusions of type B we consult
with Eq.(\ref{fuseB}) and find that the leading order contribution
with respect to the small distance $r$ is
\begin{equation}
\bbox\nabla_1\dots\bbox\nabla_q \B.F_n =
\bbox\nabla_1\dots\bbox\nabla_q 
\tilde \B.S_q(\B.r_0|\B.r_1,\B.r_2,
\dots,\B.r_q) \B.\Psi_{n,q} \ . \label{fuseBder}
\end{equation}
The contributions arising from the terms $\B.F^{(j)}_n$ all vanish
under the derivatives. To see that this is so for $2\le j\le q$ we
recall (cf. (\ref{fuseBp2}),(\ref{fuseBp3})) that although the
functions $\B.F^{(j)}_n$ depend on all the $p$ separations
$\B.r_i-\B.r_0$, it is a sum of functions $\tilde S_j$ each of which
depends only on a subset of $j$ small separations.  Thus we find a
nonzero contribution for the $q$-order derivative only from terms with
$j\ge q$.  The leading contribution for $r\ll R$ always comes from the
$j=q$ term. It is noteworthy that this contribution is independent of
the remaining $p-q$ small separations, if they exist.
  
In summary, the rule is that when we fuse $p$ coordinates on which $q$
gradients are applied, the correlation function scales asymptotically
like $r^{\zeta_p-q}R^{\zeta_n-\zeta_p}$ for fusions of type A and
$r^{\zeta_q-q}R^{\zeta_n-\zeta_q}$ for fusions of type B. The second
result is independent of the number of additional points in the ball
of size $r$ that do not have a gradient applied to them. These
additional points can be even fused together or with $\B.r_0$. This
result will be the starting point for the discussion of the
correlation function $\B.{\cal J}_n$ of the balance equation.
\section{The fusion of two points: tensor structure and 
the effects of anisotropy}

In the previous section we focused on the scaling exponents that
characterize the leading contribution to the correlation function in
the asymptotic regime when some points fused together. In this section
we address the tensor structure of the correlation functions, and the
subleading terms that exist because of the dependence on the angles
between the small separation vector and the remaining large separation
vectors. This second subject is related to the rich and important
issue of the decay of the effects of anisotropic forcing, and we do
not exhaust this issue in the present section. Some of the results are
relevant however in a much broader context.
\subsection{Tensor structure in the fusion of two points}
\label{two}
In all the previous discussion of the fusion rules when two points
(say $\B.r_1$ and $\B.r_2$) were fused we focused only on the scaling
exponent of the function $\tilde \B.F_2(\B.\rho)$ with $\B.\rho=
\B.r_1-\B.r_2$. Here we will go further in examining the structure of
the resulting correlation functions.

The fusion of two points involves just one small separation distance
$\B.\rho= \B.r_1-\B.r_2$. In general the asymptotic behavior of the
correlation function may depend on the angle between $\B.\rho$ and the
remaining large separation vectors. This dependence is discussed in
the next subsection. Here we consider the isotropic part which will be
shown to be the leading contribution.  It is easy to determine the
dependence of $\tilde \B.F_2$ on the direction of $\B.\rho$ using the
general requirement of incompressibility:
\begin{equation}
  \label{tildeS}
   \tilde F_2^{\alpha\beta}(\B.\rho)= 
A \Big[(2+\zeta_2)\delta_{\alpha\beta} 
- \zeta_2{\rho_\alpha \rho_\beta\over \rho^2}\Big]\rho^{\zeta_2}\ .
\end{equation}
This form is standard for the second-order structure function in
isotropic turbulence. We reiterate that when we extract $\tilde
\B.F_2$ out of a many-point correlation function in the process of
fusion, there is the issue of the direction of $\B.\rho$ with respect
to other vector separations which we address next.
\subsubsection{Subleading contributions: 
  effect of helicity and anisotropy}\label{twoani} 
The dependence on the angle of $\B.\rho$ is an interesting subject
that deserves full analysis in a separate study. Here we only touch on
some of the essential issues.

The analysis of these terms depends very much on the nature of the
two-point scalar correlation function $F_2(\B.\rho)\equiv
F^{\alpha\alpha}_2(\B.\rho)$ in anisotropic turbulence.  When the
forcing of turbulence is isotropic, this function depends on the
magnitude $|\B.\rho|$ only. In general however the dependence on the
orientation of $\B.\rho$ with respect to the anisotropic forces may be
important.  It is useful therefore to represent $F_2(\B.\rho)$ as a
``multipole" expansion according to
\begin{eqnarray}
  \label{multip3}
  F_2(\B.\rho)&=&\sum_{\ell=0}^\infty F_{2,\ell} 
(\B.\rho)\ , \\
F_{2,\ell} (\B.\rho)
&=&\sum_{m=-\ell}^{\ell}Y_{\ell m}(\hat \B.\rho)
\int  F_{2} (\rho\B.\xi) Y_{\ell m}(\hat \B.\xi) d \hat \B.\xi\ .
\end{eqnarray}
In this expansion the $\hat \B.\rho \equiv \B.\rho /\rho$,
$\hat\B.\xi$ is unit vector and the functions $Y_{\ell m}$ are the
standard spherical harmonics.  We chose to expand in these functions
since the relevant symmetry group in our problem is the group of
rotations SO(3).  In a scale invariant situation we expect that each
component $F_{2,\ell}$ scales like
\begin{equation}
F_{2,\ell} \propto \rho ^{\beta_\ell}\ , \label{betal}
\end{equation}
and in general the exponents $\beta_\ell$ depend on $\ell$. That this
is so with universal exponents $\beta_l$ was proved \cite{96FGLP} in
the case of Kraichnan's model of passive scalar \cite{65Kra}, but
there is yet no analogous proof in the case of Navier-Stokes
turbulence.  We will assume, in order to proceed, that the exponents
$\beta_\ell$ exist and that they are universal.

Under these assumption the calculation of the subleading contributions
to the fusion rules in the case of the fusion of two points is
straightforward.  We first consider the partial trace
$\B.F^{\alpha\alpha}_n$ of the $n$-rank tensor $\B.F_n$ with respect
to the first two indices. Next we decompose it into spherical
harmonics according to the ``multipole'' expansion\cite{96LPP}:
\begin{eqnarray}
  \label{multip1}
&&\B.F^{\alpha\alpha}_n(\B.r_0+\case{1}{2}\B.
\rho,\B.r_0-\case{1}{2}\B.\rho;
\{\B.r_k,\B.r'_k\})\nonumber \\
&=&\sum_{\ell=0}^\infty \B.F^{\alpha\alpha}_{n,\ell} 
(\B.r_0+\case{1}{2}\B.\rho,\B.r_0-\case{1}{2}\B.\rho;
\{\B.r_k,\B.r'_k\}) \,,\\
\label{multip2}
 &&\B.F^{\alpha\alpha}_{n,\ell} 
(\B.r_0+\case{1}{2}\B.\rho,\B.r_0-\case{1}{2}\B.\rho;
\{\B.r_k,\B.r'_k\})
=\sum_{m=-\ell}^{\ell}Y_{\ell m}(\hat \B.\rho)\nonumber\\
&\times&\int  \B.F^{\alpha\alpha}_{n} 
(\B.r_0+\case{1}{2}\rho\B.\xi,\B.r_0-\case{1}{2}\B.\rho\B.\xi;
\{\B.r_k,\B.r'_k\})
Y_{\ell m}
(\hat \B.\xi) d \hat \B.\xi\,,
\end{eqnarray}
The first term, $\B.F^{\alpha\alpha}_{n,0}$, corresponds to the
leading contribution with the scaling behavior ~$\rho^\zeta_2
R^{(\zeta_n-\zeta_2)}$ which was discussed above.  (Remember that $R$
is the characteristic separation in the correlation function
$\B.F^{\alpha\alpha}_n$).  The next order contributions are given by
$\B.F^{\alpha\alpha}_{n,\ell>0}$
\begin{eqnarray} \label{helic0}
&&\B.F^{\alpha\alpha}_{n,\ell}({\B.r}_1,{\B.r}'_1;{\B.r}_2,{\B.r}'_2;
\{{\B.r}_k,{\B.r}'_k\})\nonumber \\&&
=\tilde F_{2,\ell}(\B.\rho)
\bbox{\Psi}_{n,2,\ell}({\B.r}_0,;{\B.r}'_1,{\B.r}'_2;
\{{\B.r}_k,{\B.r}'_k\})\ .
\end{eqnarray}
The scaling behavior of $\tilde F_{2,\ell}$ can be read from
(\ref{betal}) under the usual assumption of universality; the two
points that fuse together relate to the unfused points in the same way
that the two-point correlation function relates to the anisotropic
forcing. We can thus write with impunity:
\begin{equation}
  \label{helic1}
\tilde F_{2,\ell} \propto \rho ^{\beta_\ell}\,,
\qquad
 \bbox{\Psi}_{n,2,\ell}\propto R^{\zeta_n - \beta_\ell}\ . 
\label{Fpsi} 
\end{equation}

To estimate the value of $\beta_1$ and $\beta_2$ we need to understand
what is the physics that determines them.  In fact, since we
considered the partial trace of the correlation function, we end up
with $\tilde F_2$ which is even under the transformation $\B.\rho \to
-\B.\rho$. Accordingly, although the exponents are as stated in
(\ref{Fpsi}) the coefficients of all odd $\ell$ quantities are zero.
In order to examine odd $\ell $ contributions one needs to form a
correlator which is not even in $\B.\rho$. As an example we consider
\begin{equation}
  \label{nonsym}
F_2^{\alpha\beta}(\B.\rho)=\big\langle u_\alpha (\B.r+\B.\rho)
u_\beta (\B.r)-  u_\alpha (\B.r-\B.\rho)
u_\beta (\B.r)\big\rangle \ .
\end{equation}
Since this object is manifestly odd in $\B.\rho$ it vanishes when
there exists inversion symmetry. For turbulence with non-zero helicity
~$F_2^{\alpha\beta}(\B.\rho)\ne 0$ and the leading contribution to
~$F_2^{\alpha\beta}(\B.\rho)$ (which is
~$F_{2,1}^{\alpha\beta}(\B.\rho)$) is determined by the flux of
helicity. This is reminiscent of the flux of energy which determines
the leading contribution to the second order structure function.
Standard K41 dimensional reasoning leads to the prediction
$\beta_1=1$, see for example\cite{81KL}. This holds probably with the
same accuracy as the K41 prediction for $\zeta_2$, which is
$\zeta_2=2/3$ instead of the experimental value $\zeta_2\approx 0.70$
\cite{93SK,93Ben}.

We can easily determine the tensor structure of
~$F_{2,1}^{\alpha\beta}(\B.\rho)$ in isotropic incompressible
turbulence (in the absence of inversion symmetry):
\begin{equation}
  \label{helicity}
 F_{2,1}^{\alpha\beta}(\B.\rho)=\epsilon_{\alpha\beta\gamma}
 \hat \rho_\gamma F_{2,1} (\rho) \, , \quad 
F_{2,1} (\rho)\propto \rho^{\beta_1}\,,
 \end{equation}
 where $\epsilon_{\alpha\beta\gamma}$ isq the fully antisymmetric unit
 tensor ($\epsilon_{123}=-\epsilon_{213}=1 $). As we mentioned, in the
 presence of inversion symmetry $\B.F_2(-\B.\rho)=\B.F_2(\B.\rho)$ and
 all terms which are odd in $\ell$ in (\ref{multip3}) are zero. On the
 other hand this is not the case for $\tilde\B.F_{2\ell}(\B.\rho)$,
 which appears in the fusion of two points in a many-point correlation
 function.  Even if the turbulent flow itself has inversion symmetry,
 the geometry of all the points appearing in (\ref{helic0}) can lead
 to non-vanishing odd $\ell$ components of $\tilde\B.F_{2\ell}$. The
 positions of the points $\B.r'_1,\B.r'_2\dots$ are such that there is
 no inversion symmetry around the center of the fusing coordinates
 $\B.r_1$ and $\B.r_2$ which is $\B.r_0=(\B.r_1+\B.r_2)/2$.  Therefore
 ~$\tilde\B.F_{2\ell}(\B.\rho)\ne 0$ and because of the same
 constraints it has the same tensor structure as (\ref{helicity}):
\begin{equation}
  \label{helicity2}
 \tilde F_{2,1}^{\alpha\beta}(\B.\rho)=\epsilon_{\alpha\beta\gamma}
 \hat \rho_\gamma \tilde F_{2,1} (\rho) \, , \quad 
\tilde F_{2,1} (\rho)\propto \rho^{\beta_1}\ . 
 \end{equation}
 The physical origin of this term in the multipole expansion is a local
 flux of helicity. Even when the average helicity flux is zero,
the local value of the flux conditioned on the velocities fixed at
certain coordinates may be non zero. 

The exponent $\beta_2$ is the leading exponent describing the rate of
decay of the effects of anisotropy, and may be computed using
perturbation theory, disregarding the nonperturbative effects which
are the subject of this paper, see \cite{94GLLP,95FL} and reference
therein. The result is $\beta_2=4/3$, and again one expects this
result to be numerically close to the truth. We do not possess
presently any numerical estimates for $\beta_\ell$ with higher values
of $\ell$, and as we said before it is not guaranteed that these
exponents are universal. These issues have to be considered
independently in the future.
\section{Generalized Balance Equation}
\label{sec:GBE}
\subsection{Derivation of generalized balance equation}
The starting point of this analysis is the Navier-Stokes equations
for incompressible flows:
\begin{eqnarray}
{\partial \B.u (\B.r,t) \over \partial t} + \B.u (\B.r,t)\cdot
\bbox\nabla \B.u
(\B.r,t)+\bbox\nabla p(\B.r,t) &=&\nu \nabla^2\B.u (\B.r,t) \ ,
\nonumber \\
\bbox\nabla\cdot \B.u(\B.r,t)&=&0 \ . \label{NS}
 \end{eqnarray}
 In general we need to add a forcing term to these equations. It was
 shown in \cite{LP-3} that as far as the balance equations are
 concerned, the effect of the forcing term is felt in the energy
 containing scales only. For this reason we do not write the forcing
 explicitly.  As usual the gradient of the pressure in (\ref{NS}) is
 eliminated by applying the transverse projection operator $\tensor P$. The
 Navier-Stokes equations takes on the form
 \begin{equation}
 {\partial{\B.u}({\B.r},t)\over\partial t} 
+{\tensor P}\left [{\B.u}({\bf r},t)\cdot\bbox\nabla
\right]{\B.u}({\B.r},t)=\nu\nabla^2{\B.u}({\B.r},t)\ .
\label{NSpro}
 \end{equation}
 The application of $\tensor P$ 
to any given vector field ${\B.a}( {\B.r})$ is
 non-local, and has the form:
 \begin{equation}
 {\tensor P} {\B.a}(
 {\B.r})\rbrack_\alpha =\int d   {\B.r}'   P_{\alpha\beta}(
 \B.r')a_\beta({\B.r}-{\B.r}'),
 \label{b2}
 \end{equation}
 where the kernel $P_{\alpha\beta}({\B.r})$ is the following
 difference:
 \begin{equation}\label{b5}
 P_{\alpha\beta}({\B.r})
 =\delta_{\alpha\beta}\delta({\B.r} )
-P_{\alpha\beta}^{||}({\B.r})\ .
\end{equation}
Here $P_{\alpha\beta}^{||}({\B.r})$ is the kernel of the longitudinal
projector which appears here due to the effect of the pressure term in
the Navier-Stokes equation:
\begin{equation}\label{long}
P_{\alpha\beta}^{||}({\B.r}) 
={1\over 4\pi}\left[{\delta_{\alpha\beta}
 \over r^3}
 -{3r_\alpha r_\beta
 \over  r^5}\right] \ .
 \end{equation}
 Given the equation of motion we can take the time derivative of
 Eq.(\ref{defF}). We find
\begin{eqnarray}
{\partial \B.F_n\over \partial t} &=&\sum_{j=1}^n
\Big\langle\B.w(\B.r_1,\B.r'_1,t)\dots
\label{dtFn}\\
&&\dots
{\partial\B.w(\B.r_j,\B.r'_j,t)\over \partial t}
\dots \B.w(\B.r_n,\B.r'_n,t)\Big\rangle
\ .\nonumber 
\end{eqnarray}
Substituting Eq.(\ref{NSpro}), and considering the stationary state in
which $\partial \B.F_n/ \partial t=0$ we find the balance equations
\begin{equation}
\B.{\cal D}_n(\B.r_1,\B.r'_1;\dots \B.r_n,\B.r'_n) 
=\B.{\cal J}_n(\B.r_1,\B.r'_1;\dots
\B.r_n,\B.r'_n)
\ , \label{bal}
\end{equation}
where the ``interaction'' term $\B.{\cal D}_n$ stems from the
nonlinear and pressure terms:
\begin{eqnarray}
&&\B.{\cal D}_n(\B.r_1,\B.r'_1;\dots \B.r_n,\B.r'_n)=\sum_{j=1}^n
\langle\B.w(\B.r_1,\B.r'_1)\dots
\label{Dn} \\
 &\dots&\left[\left({\tensor P} {\B.u}\cdot\bbox\nabla {\B.u}\right)_j-
\left({\tensor P} {\B.u}\cdot\bbox\nabla {\B.u}\right)_{j'}\right]
\dots \B.w(\B.r_n,\B.r'_n)\rangle\ ,
\nonumber 
\end{eqnarray}
and the ``dissipative'' term $\B.{\cal J}_n$ originates from the 
viscosity term in the Navier Stokes equation:
\begin{eqnarray}
&&\B.{\cal J}_n(\B.r_1,\B.r'_1;\dots
\B.r_n,\B.r'_n)=\nu\sum_{j=1}^n\left(\nabla_j^2+\nabla_{j'}^2\right)
\langle\B.w(\B.r_1,\B.r'_1)\dots \nonumber \\ &&\dots
\B.w(\B.r_j,\B.r'_j)
\dots \B.w(\B.r_n,\B.r'_n)\rangle \ . \label{Jn}
\end{eqnarray}
In writing these equations we used the fact that in the stationary
state the time designation is unneeded. We also used the following
short-hand notation:
\begin{equation}
\left({\tensor P} {\B.u}\cdot\bbox\nabla {\B.u}\right )^\alpha_j=
\int{d\B.r}P_{\alpha\beta}(\B.r_j-\B.r)u_\gamma(\B.r)\nabla_\gamma
u_\beta(\B.r) \ , \label{sh}
\end{equation}
and  denoted by $\nabla^2_j$ the Laplacian operator acting on
$\B.r_j$.  Equation (\ref{Dn}) can be written explicitly in the form
\begin{eqnarray}
&&{\cal D}^{\alpha_1\alpha_2\dots \alpha_n}_n(\B.r_1,\B.r'_1;\dots
\B.r_n,\B.r'_n)=
\sum_{j=1}^n \int d\B.r P_{\alpha_j\beta}(\B.r)
\label{yak}\\
&\times&\langle w_{\alpha_1}(\B.r_1,\B.r'_1)
\dots L_\beta(\B.r_j,\B.r'_j,\B.r)
\dots w_{\alpha_n}(\B.r_n,\B.r'_n)\rangle\,,
\nonumber
\end{eqnarray}
where
\begin{eqnarray}
L_\beta(\B.r_j,\B.r'_j,\B.r)&\equiv&
\Big[\B.u(\B.r_j-\B.r)\cdot{\bbox\nabla}_j
u_\beta(\B.r_j-\B.r)
\label{Lb}\\
&&-\B.u(\B.r'_j-\B.r)\cdot{\bbox\nabla}'_j
u_\beta(\B.r'_j-\B.r) \Big]\ .
\nonumber
\end{eqnarray}
Now we begin to analyze the balance equation 
(\ref{bal}, \ref{Jn}, \ref{yak}, \ref{Lb}).
\subsection{Galilean invariance  of the generalized balance equation}
\label{derGBE}
The balance equation must be Galilean invariant. Equation (\ref{Jn})
for $\B.{\cal J}_n$ depends only on velocity differences and clearly
is Galilean invariant. This is not so obvious in the case of Eq.
(\ref{yak}) for $\B.{\cal D}_n$ because $L_\beta $ in (\ref{Lb})
contains velocities itself.  In order to clarify the Galilean
invariance of $\B.{\cal D}_n$ let us express $L_\beta$ via velocity
differences only. The first step is to subtract from $u_\beta
(\B.r_j-\B.r)$ under the derivative ${\bbox\nabla}_j$ the velocity
$u_\beta (\B.r'_j-\B.r)$ (independent of $\B.r_j$) and from $u_\beta
(\B.r'_j-\B.r)$ under the derivative ${\bbox\nabla}'_j$ the velocity
$u_\beta (\B.r_j-\B.r)$ (independent of $\B.r'_j$).  Then Eq.
(\ref{Lb}) takes the form
\begin{eqnarray}
\label{Lb1}
L_\beta(\B.r_j,\B.r'_j,\B.r)&=&
\big[\B.u(\B.r_j-\B.r)\cdot{\bbox\nabla}_j
+ \B.u(\B.r'_j-\B.r)\cdot{\bbox\nabla}'_j \big]\\
&&\times w_\beta(\B.r_j-\B.r,\B.r'_j-\B.r)\ .
\nonumber
\end{eqnarray}
For the next step let us introduce
$\bar\B.u(\B.r_1,\B.r'_1;\dots;\B.r_n,\B.r'_n)$ as the mean velocity
over all $2n$ space coordinates:
\begin{equation}
  \label{meanvel}
  \bar \B.u \equiv {1\over 2n}\sum_{k=1}^n \big[\B.u(\B.r_k)
+\B.u(\B.r'_k)\big] 
\end{equation}
and the following velocity differences:
\begin{equation}
  \label{veldif}
  \B.w(\B.r_j)\equiv \B.u(\B.r_j)-\bar \B.u\,, \quad 
 \B.w(\B.r'_j)\equiv \B.u(\B.r'_j)-\bar \B.u\ .
\end{equation}
For brevity we do not display here (and below) arguments of the
velocity $\bar \B.u$. Using (\ref{veldif}) we can present $L_\beta$
in (\ref{Lb}) as a sum of two terms:
\begin{equation}
  \label{Lb2}
  L_\beta(\B.r_j,\B.r'_j,\B.r)=L_\beta^{(1)}(\B.r_j,\B.r'_j,\B.r)
+L_\beta^{(2)}(\B.r_j,\B.r'_j,\B.r)\ .
\end{equation}
Here the first  term depends only on velocity differences:
\begin{eqnarray}
 L_\beta^{(1)}(\B.r_j,\B.r'_j,\B.r)&=&\big[\B.w(\B.r_j-\B.r)
\cdot{\bbox\nabla}_j
\label{Lb3}\\
&&+ \B.w(\B.r'_j-\B.r)\cdot{\bbox\nabla}'_j \big]
w_\beta(\B.r_j-\B.r,\B.r'_j-\B.r)\ .
\nonumber
\end{eqnarray}
However the second term does not have this property:
\begin{equation}
 \label{Lb4}
 L_\beta^{(2)}(\B.r_j,\B.r'_j,\B.r)=\bar \B.u\cdot \big[
{\bbox\nabla}_j +\cdot{\bbox\nabla}'_j \big]
w_\beta(\B.r_j-\B.r,\B.r'_j-\B.r)
\end{equation}
but, as we are going to show, this term gives zero contribution to Eq.
(\ref{yak}) for $\B.{\cal D}_n$. Indeed, by substituting (\ref{Lb4})
in (\ref{yak}) we have:
\begin{eqnarray} 
&&\B.{\cal D}_n^{(2)}=
\sum_{j=1}^n \int d\B.r_j P_{\alpha_j\beta}(\B.r)
\label{yak2}\\
&\times&\langle w_{\alpha_1}(\B.r_1,\B.r'_1)
\dots L^\beta_2(\B.r_j,\B.r'_j,\B.r)
\dots w_{\alpha_n}(\B.r_n,\B.r'_n)\rangle\ .
\nonumber
\end{eqnarray}
In its turn, this equation may be written as the difference of two
terms, $\B.{\cal D}_n^{(2)}=\B.{\cal D}_n^{(2a)}-\B.{\cal
  D}_n^{(2b)}$, which correspond to the two terms in the
Eq.~(\ref{b5}) for the kernel ~$P_{\alpha_j\beta}(\B.r)$. By
substituting in (\ref{yak2})
~$P_{\alpha\beta}(\B.r)=\delta_{\alpha\beta}\delta(\B.r)$ we have:
 \FL
\begin{eqnarray}
\!\B.{\cal D}_n^{(2a)}&\!=\!&
\sum_{j=1}^n \langle w_{\alpha_1}(\B.r_1,\B.r'_1)
\!\dots \!L^\beta_2(\B.r_j,\B.r'_j,0)
\!\dots \!w_{\alpha_n}(\B.r_n,\B.r'_n)\rangle.
\nonumber\\
&& \label{D21}
\end{eqnarray}
Using the longitudinal projector $P_{\alpha_j\beta}^{||}(\B.r)$
instead of the transversal one $P_{\alpha_j\beta}(\B.r)$ in Eq.
(\ref{yak2}) we have:
\begin{eqnarray}\label{D22}
\B.{\cal D}_n^{(2b)}&=&
\sum_{j=1}^n \int d\B.r_j P_{\alpha_j\beta}^{||}(\B.r)
\\
&\times&\langle w_{\alpha_1}(\B.r_1,\B.r'_1)
\dots L_\beta^{(2)}(\B.r_j,\B.r'_j,\B.r)
\dots w_{\alpha_n}(\B.r_n,\B.r'_n)\rangle\ .
\nonumber
\end{eqnarray}

Let us show that both these terms are zero (but because of different
reasons).  Consider the first expression for $\B.{\cal D}_n^{(2a)}$.
Substituting the explicit form (\ref{Lb4}) for
$L_\beta^{(2)}(\B.r_j,\B.r'_j,0)$ and using the incompressibility
constraint (which allows one to commute ${\bbox\nabla}_j
+{\bbox\nabla}'_j$ and $ \bar \B.u$) one has:
\begin{eqnarray}
\B.{\cal D}_n^{(2a)}&=&
\sum_{j=1}^n \Big\langle w_{\alpha_1}(\B.r_1,\B.r'_1)
\label{D21a}\\
\dots &\{\big[( &
{\bbox\nabla}_j +{\bbox\nabla}'_j )\cdot  \bar \B.u \big]
w_{\alpha_j}(\B.r_j,\B.r'_j)  \}
\dots w_{\alpha_n}(\B.r_n,\B.r'_n)\Big\rangle .
\nonumber
\end{eqnarray} 
This equation may be rewritten as
\begin{eqnarray}
&&\B.{\cal D}_n^{(2a)}=
\sum_{j=1}^n 
\big(\nabla^\beta_j +\nabla'^\beta_j\big)
\label{D21b}\\
&\times&\Big\langle \bar u^\beta
w_{\alpha_1}(\B.r_1,\B.r'_1)
w_{\alpha_2}(\B.r_j,\B.r'_2) 
\dots w_{\alpha_n}(\B.r_n,\B.r'_n)\Big\rangle \ .
\nonumber
\end{eqnarray}
Remember that due to space homogeneity the correlation function in the
second line of this equation is independent of the sum of coordinates
~$\sum_{j=1}^n (\B.r_j+\B.r'_j)$. Therefore $\B.{\cal D}_n^{(2a)}$ is
indeed equal to zero. 

Consider next Eq. (\ref{D22}) for ~$\B.{\cal D}_n^{(2b)}$. According
to (\ref{Lb4}) $L_\beta^{(2)}\propto w_\beta (\B.r_j-\B.r,
\B.r'_j-\B.r)$ and acting on this velocity with the longitudinal
projector gives zero because of the incompressibility constraint. Thus
we can conclude that $\B.{\cal D}_n^{(2a)}= \B.{\cal D}_n^{(2b)}=0$.
Therefore we can get an expression for $\B.{\cal D}_n$ by replacing $
L_\beta $ in Eq.  (\ref{yak}) with $ L_\beta^{(1)}$ taken from Eq.
(\ref{Lb3}):
\begin{eqnarray}
&&{\cal D}^{\alpha_1\alpha_2\dots \alpha_n}_n(\B.r_1,\B.r'_1;\dots
\B.r_n,\B.r'_n)=-\int d\B.r \sum_{j=1}^n  P_{\alpha_j\beta}(\B.r)
\nonumber\\ 
&\times& \Big\langle w_{\alpha_1}(\B.r_1,\B.r'_1)
\dots \Big\{ \big[ u_\gamma(\B.r_j\!-\!\B.r)
-\bar u_\gamma \big]{\partial \over \partial r_{j\gamma}}
\nonumber \\&+&\big[u_\gamma(\B.r'_j\!-\!\B.r)-\bar u_\gamma \big] 
{\partial \over \partial r'_{j\gamma}}\Big\}
\nonumber \\
&\times&
w_\beta (\B.r_j-\B.r,\B.r'_j-\B.r)
\dots w_{\alpha_n}(\B.r_n,\B.r'_n)\Big \rangle\ .
\label{Dnfin}
\end{eqnarray}
This expression for $\B.{\cal D}_n$ depends only on velocity
differences and therefore the Galilean invariance becomes obvious.

\subsection{Locality of the Interaction Term}
We begin to analyze the interaction terms $\B.{\cal D}_n$
(\ref{Dnfin}) for the most general configuration for which all the
$2n$ coordinates $\B.r_j$, $\B.r'_j$ are different, and all the
$n(2n-1)$ separations are of the same order of magnitude, which we
designate by $R$.  This analysis when $R$ is in the inertial interval
follows the ideas of the analysis presented in \cite{LP-3} section 6A
for the interaction term of the structure function. The interaction
term that we face here is a significantly more complicated object. The
main point is that the integral over $\B.r$ appearing in Eq.
(\ref{Dnfin}) for $\B.{\cal D}_n$ is ``local" in the following sense.
First it converges in the ``ultra-violet" (UV) limit. This limit has
to be considered when (i) $r\to 0$, (ii) when $(\B.r_j-\B.r)$ becomes
very close to any of the $2n-1$ coordinates other than $r_j$, and
(iii) when $(\B.r'_j-\B.r)$ becomes very close to any of the $2n-1$
coordinates other than $r'_j$.  Second, it converges in the
``infra-red" (IR) limit when $r\to \infty$.  The idea for the proof of
these properties lies in the use of the fusion rules which we
discussed in Sect.~\ref{FRII}.
\subsubsection{Ultraviolet convergence}
\label{UVloc} 
To demonstrate the convergence of the integral in $\B.{\cal D}_n$ in
the ultraviolet region we can consider any term from the sum on $j$.
Writing $u_\gamma(\B.r_j -\B.r) =
(\sum_{k=1}^nu_\gamma(\B.r_j-\B.r))/n$, and using Eq.(\ref{meanvel}),
we consider one of the $k$-terms in the sum.  The integral that
appears is of the form
\begin{eqnarray}
&&I={1\over n}\int d\B.r  P_{\alpha_j\beta}(\B.r)
{\partial \over \partial r_{j\gamma}} 
 \Big\langle w_{\alpha_1}(\B.r_1,\B.r'_1)
\dots \label{term}\\
&&w_\gamma(\B.r_j\!-\!\B.r,\B.r_k)
w_\beta (\B.r_j-\B.r,\B.r'_j-\B.r)
\dots w_{\alpha_n}(\B.r_n,\B.r'_n)\Big \rangle\ .
\nonumber
\end{eqnarray}
As the coordinate $\B.r$ is being integrated over, the most dangerous
ultraviolet contribution comes from the region of small $r$. In this
region the projection operator can be evaluated as $1/r^3$. Other
coalescence events of $\B.r$ with other coordinates contribute less
divergent integrands since the projection operator does not become
singular. When $r$ becomes small, there are two possibilities: (i)
$\B.r_j \ne \B.r_k$ and (ii) $\B.r_j =\B.r_k$. In the first case the
correlation function itself is analytic in the region $r \to 0$, and
we can expand it in a Taylor series ${\rm Const}+ \B.B\cdot\B.r
+\dots$.  where $\B.B$ is an $\B.r$-independent vector. The constant
term is annihilated by the projection operator. The term linear in
$\B.r$ vanishes under the $d\B.r$ integration due to $\B.r\to -\B.r$
symmetry. The next term which is proportional to $r^2$ is convergent
in the ultraviolet. In the second case we have a velocity difference
across the length $r$. Accordingly we need to use the fusion rule
(\ref{funny}), and we learn that the leading contribution is
proportional to $r^{\zeta_2}$. This is sufficient for convergence in
the ultraviolet. We note that the derivative with respect to $r_j$
cannot be evaluated as $1/r$ when $\B.r_j=\B.r_k$. Rather, it is
evaluated as the inverse of the distance between $\B.r_j$ and the
nearest coordinate in the correlation function.
\subsubsection{Infrared convergence}
\label{IRloc}
\begin{figure}
\label{Fig3}
\end{figure}

{\small FIG.~\ref{Fig3}. Typical geometry with $(n-1)$ velocity
  differences in a ball of radius $R$ on the left separated by a large
  distance $r\gg R$ from a pair of points on the right.}\vskip .4cm

To understand the convergence of $\B.{\cal D}_n$ when the integration
variable $r$ becomes very large we can consider again the typical term
(\ref{term}).  The relevant geometry is shown in Fig.~\ref{Fig3}.
There is one velocity difference across the coordinates $\B.r_j-\B.r$
and $\B.r'_j-\B.r$ (which is shown on the right of the figure),
$(n-1)$ velocity differences across coordinates that are all within a
ball of radius $R$ (at the left of the figure), and one velocity
difference across the large distance $r$ which is much larger than
$R$. In the notation of this figure the leading order contribution for
large $r$ is obtained from the fusion rules (\ref{funny}) for the
situation on the right and (\ref{fusion1}) for the geometry on the
left. The resulting evaluation for the leading term is
\begin{equation}
I \propto r^{\zeta_{n+1}}\left({R_j\over r}\right)^{\zeta_2}
\left({R\over r}\right)^{\zeta_{n-1}} \ . \label{noworry}
\end{equation}
On the face of it, this term is near dangerous. For K41 scaling the
$r$-dependence cancels, and the integral is logarithmically divergent.
For anomalous scaling the integral converges since $\zeta_{n+1}\le
\zeta_{n-1}+\zeta_2$ due to Hoelder inequalities. This convergence
seems slow. However, the situation is in fact much safer.  If we take
into account the precise form of the second-order structure function
in the fusion rules we find that the divergence with respect to
$\B.r_j$ translates in fact to ${\partial S_2^{\beta
    \gamma}(\B.R_j)/\partial R_{j\gamma}}$ which is zero due to
incompressibility.  The next order term is convergent even for simple
(K41) scaling. This completes the proof of locality of (\ref{Dn}). The
conclusion is that when all the separations in $\B.D_n$ are of the
same order of magnitude $R$, the main contribution to the integral in
(\ref{Dn}) comes from the region $r\sim R$.  Therefore, the integral
can be evaluated by straightforward power counting, leading to
\begin{equation}
{\cal D}_n\sim {S_{n+1}(R)\over R} \ . \label{Deval} 
\end{equation}
It should be stressed that a more detailed analysis demonstrates that
when the separations $\B.\rho$ between the coordinates that do not
involve velocity differences, (i.e separations like $\B.r_{jk}$ but
not $\B.R_j$) go to zero, the evaluation does not change. A direct
proof of this fact for the case when all such separations are fused
(i.e the standard structure function) was presented in \cite{LP-3}. On
the other hand, if we consider fusions of type A, in which coordinates
across a velocity difference coalesce ($\B.\rho=\B.R_j$), we need to
be more careful. When two points undergo a fusion of type A the rough
evaluation of $\B.{\cal D}_n$ is
$[(dS_2(\rho)/d\rho)][S_{n+1}(R)/S_2(R)]$ where $R$ is the
characteristic value of large separations. However taking into account
the tensor structure of the first factor one sees that it vanishes due
to the incompressibility constraint. In the next-order term the
evaluation of the gradient is $1/R$ and $\B.{\cal D}_n\to 0$ when
$\rho\to 0$.

\subsection{The dissipative term}
The evaluation of the quantity $\B.{\cal J}_n$ is more
straightforward. When all the separations $R_j$ and $r_{ij}$ are of
the same order $R$, the correlator in (\ref{Jn}) is evaluated simply
as $S_n(R)$. The Laplacian is then of the order of $1/R^2$. The
evaluation is
\begin{equation}
{\cal J}_n \sim \nu {S_n(R)\over R^2} \ . \label{JnR}
\end{equation}
When one of the separations becomes much smaller than the rest the
evaluation can be read directly from the definition (59) and from the
fusion rules. Denoting the smallest separation by $r_{min}$ we write
\begin{equation}
{\cal J}_n \sim \nu S_2(r_{\rm min}){S_n(R)\over S_2(R)r_{\rm min}^2} \ . 
\label{JnRmin}
\end{equation}
\section{The homogeneous equation} 
We noted that when all the separations involved in our correlation
functions are of the same order of magnitude, and when $\nu \to 0$
(which is the limit of infinite Reynolds number Re), the term ${\cal
  J}_n$ becomes negligible compared to ${\cal D}_n$. The ratio ${\cal
  J}_n/{\cal D}_n$ is evaluated as $\nu S_n(R)/RS_{n+1}(R)$, which for
fixed $R$ vanishes in the limit $\nu \to 0$.  Thus the ``balance
equation" becomes a {\em homogeneous} integro-differential equation
$\B.{\cal D}_n =0$ which may have scale-invariant solutions with
anomalous scaling exponents $\zeta_{n+1}\ne (n+1)/3$. It should be
stressed that the evaluation (\ref{Deval}) remains correct for every
term in $\B.{\cal D}_n$, but various terms cancel to give zero in the
homogeneous equation, {\em provided that the scaling exponent
  $\zeta_n$ is chosen correctly}. To make this important point clear
we exemplify it with the simple case $n=2$ for which $\B.{\cal D}_n$
can be greatly simplified.  Consider the scalar object ${\cal
  F}_2(\B.r_1,\B.r'_1,\B.r_2, \B.r'_2) \!=\!
\left<\B.w(\B.r_1|\B.r'_1)\cdot \B.w(\B.r_2|\B.r'_2)\right>$. The
terms in the scalar balance equation for this case are exactly
\begin{eqnarray}
{\cal D}_2(\B.r_1,\B.r'_1,\B.r_2,\B.r'_2)&=&d [S_3(r_{12'})
-S_3(r_{12})]/2dr_1\nonumber \\
&+&d[S_3(r_{1'2})-S_3(r_{1'2'})]/2dr'_1
\ , \label{D2}\\
{\cal J}_2(\B.r_1,\B.r'_1,\B.r_{2},\B.r'_{2})
&=&\nu\{\nabla^2_1[S_2(r_{12'})
-S_2(r_{12})]\nonumber \\
&+&\nabla^2_{1'}[S_2(r_{1'2})-S_2(r_{1'2'})]\} \ . \label{J2}
 \end{eqnarray}
 When all the separations are of the order of $R$ we can see
 explicitly that ${\cal J}_2\sim \nu S_2(R)/R^2$ which is much smaller
 than each term in ${\cal D}_2$. Considering the scale invariant
 solution $S_3(R) =A R^{\zeta_3}$ where $A$ is a dimensional
 coefficient, we see that
$$
{\cal D}_2(\B.r_1,\B.r'_1,\B.r_{2},\B.r'_{2})={\zeta_3 A\over 2}\Big[
r_{12'}^{\zeta_3-1}-r_{12}^{\zeta_3-1}+r_{1'2}
^{\zeta_3-1}-r_{1'2'}^{\zeta_3-1}\Big].
$$
Obviously the solution for ${\cal D}_2=0$ requires the unique
choice $\zeta_3=1$ which is the known exponent for $S_3$ \cite{Fri}.
The coefficient $A$ is now determined as $\bar\epsilon$ which is the
mean energy dissipation per unit mass and unit time.

We presently do not know how to find the homogeneous solution of the
equation $\B.{\cal D}_n=0$ for higher values of $n$. We are even not
fully confident that this is an equation in the usual sense and not
just a constraint that will not be sufficient for a unique
determination of the scaling exponents $\zeta_n$. We feel however that
this is an interesting equation that will offer interesting and
worthwhile insights.
\section{The dissipative scaling functions}
In this section we consider the dissipative ``scales" and show that
they are actually scaling functions. To define properly the
dissipative length we use the fact that there is a cross-over from the
scale invariant solution of the homogeneous equation to dissipative
solutions when ${\cal J}_2$ becomes comparable to any of the terms in
${\cal D}_2$. This happens when at least one of the separations
appearing in (\ref{J2}) becomes small enough. Denoting the smallest
separation as $r_{\rm min}$ we evaluate ${\cal J}_2\sim \nu S_2(r_{\rm
  min})/r_{\rm min}^2$. From this we can estimate, using the balance
equation, $S_2(r_{\rm min})\sim (S_3(R)/\nu R)r_{\rm min}^2\sim \bar
\epsilon r_{\rm min}^2/\nu$.  In the inertial range we have $S_2(r)
\sim (\bar\epsilon r)^{2/3}(r/L)^{\zeta_2-2/3}$.  The viscous scale
$\eta_2$ for the second-order structure function is then determined
from finding where these two expressions are of the same order of
magnitude, i.e.
\begin{equation}
\bar\epsilon \eta_2^2/\nu 
= (\bar\epsilon \eta_2)^{2/3}(r/L)^{\zeta_2-2/3}. \label{defeta}
\end{equation}
Using the outer velocity scale $U_L$ we estimate $\bar\epsilon \sim
U_L^3/L$ and end up with
\begin{equation}
\eta_2 \sim L{\rm Re}^{-1/(2-\zeta_2)} \ . \label{eta2}
\end{equation}
Note that this result is not in agreement with the ad-hoc application
of the multifractal model \cite{Fri,87PV,91FV} which predicts $\eta_2
\sim L{\rm Re}^{-2/(2+\zeta_2)}$.

A similar mechanism operates in the general case of $n\ne 2$. As long
as all the separations are in the inertial interval $\B.{\cal J}_n$ is
negligible.  When one separation e.g. $r_{12}$ diminishes towards
zero, and all the other separations are of the order of $R$, the
internal cancellations leading to the homogeneous equation $\B.{\cal
  D}_n=0$ disappear, and $\B.{\cal D}_n$ is evaluated as in
(\ref{Deval}).  The term $\B.{\cal J}_n$ is now dominated by one
contribution that can be written in short-hand notation as
$\nu\nabla^2_1F_n(r_{12},\{R\})$. We can solve for $F_n(r_{12},\{R\})$
in this limit:
\begin{equation}
F_n(r_{12},\{R\})\approx r_{12}^2 S_{n+1}(R)/ \nu R \ .\label{dissFn}
\end{equation}
On the other hand we have, from the fusion rule (\ref{fusion2}), the
form of the same quantity when $r_{12}$ is still in the inertial
range, i.e.  $F_n(r_{12},\{R\}) \approx S_2(r_{12})S_n(R)/S_2(R)$. To
estimate the viscous scale $\eta_n$ we find when these two evaluations
are of the same order.  The answer is
\begin{equation}
\eta_n(R) = \eta_2 \Big({R\over L}\Big)^{x_n},\quad 
x_n={\zeta_n+\zeta_3-\zeta_{n+1}-\zeta_2\over
2-\zeta_2} \ . \label{finaletan}
\end{equation}
We note that the Hoelder inequalities guarantee that $x_n>0$ and
increases with $n$. We see that the viscous ``length" is actually an
anomalous scaling function.
\section{Exact bridge relations}
In this section we derive important (and exact) scaling relations
between the exponents $\zeta_n$ of the structure functions and
exponents involving correlations of the dissipation field.  We
consider correlations of the type \FL
\begin{eqnarray}
\B.{\cal K}_{\epsilon}^{(n)}&\equiv& \left<\epsilon(\B.x_1)
\B.w(\B.r_1|\B.r'_1)\!\dots \!\B.w(\B.r_n|\B.r'_n)\right>
\propto R^{-\mu^{(1)}_n},
\label{Ken}\\
\B.{\cal K}_{2\epsilon}^{(n)}&\equiv& \left<\epsilon(\B.x_1)
\epsilon(\B.x_2)
\B.w(\B.r_1|\B.r'_1)\!\dots \!\B.w(\B.r_n|\B.r'_n)\right>
\propto R^{-\mu^{(2)}_n}, 
\label{Keen}\\
\B.{\cal K}_{p\epsilon}^{(n)}&\equiv& \left<\epsilon(\B.x_1)
\epsilon(\B.x_2)\dots
\epsilon(\B.x_p)
\B.w(\B.r_1|\B.r'_1)\!\dots \!\B.w(\B.r_n|\B.r'_n)\right>\nonumber\\
&\propto& R^{-\mu^{(p)}_n}, 
\label{Kpen}
\end{eqnarray}
where $R$ is a typical separation between any pair and
$\epsilon(\B.x)\equiv \nu|\nabla \B.u(\B.x)|^2$, and we are interested
in the scaling relations between the exponents $\mu^{(p)}_n$ and the
exponents $\zeta_n$. Note that $\mu^{(2)}_0$ in this notation is the
well studied \cite{92Pra,93SK} exponent of dissipation fluctuation
which is denoted $\mu$. We begin with the rigorous calculation of
$\mu^{(1)}_n$.

$\!\!$Consider (\ref{Jn}) for $\B.{\cal
  J}_{n+2}(\B.x_1,\B.x'_1;\B.x_2,\B.x'_2;\B.r_1,\B.r'_1\dots
\B.r_n,\B.r'_n)$ in the limit $\B.x_1\to \B.x_2$. The leading
contribution in the limit arises from the Laplacians with respect to
the coalescing points:
\begin{eqnarray}
\lim_{x_1\to x_2} \B.{\cal J}_{n+2} &=& 
\nu\lim_{x_1\to x_2}(\nabla^2_1+\nabla^2_2)
\langle\B.u(\B.x_1)\B.u(\B.x_2)\nonumber \\
&\times&\B.w(\B.r_1,\B.r'_1)
\dots\B.w(\B.r_n,\B.r'_n)\rangle
\ . \label{limJn}
\end{eqnarray}
Moving one gradient around and taking the trace with respect to the
first two tensor indices we see that in this limit
\begin{equation}
\lim_{r_1\to r_2} \B.{\cal J}^{\alpha\alpha}_{n+2}
 = -2\B.{\cal K}_{\epsilon}^{(n)} \ .
\end{equation}
As explained in the previous section, when the two points coalesce
$\B.{\cal D}_{n+2}$ of the balance equation loses its internal
cancellations, and we can therefore conclude immediately that
\begin{equation}
\B.{\cal K}_{\epsilon}^{(n)}\sim {S_{n+3}(R)\over R}
\ .\label{KF}
\end{equation}
In terms of the scaling exponents we are led to the exact relation
\begin{equation}
\mu^{(1)}_n = 1-\zeta_{n+3} . \label{mu1n}
\end{equation}
The scaling relations satisfied by $\mu^{(2)}_n$ require
considerations of the second time derivative of the correlation
(\ref{defF}).  \FL
\begin{eqnarray}
\ddot{\bbox{\cal F}}_n=\sum_{i,j=1}^n
\langle\B.w(\B.r_1|\B.r'_1,t) \dots\dot\B.w(\B.r_i|\B.r'_i,t)
\nonumber\\
\dots\dot\B.w(\B.r_j|\B.r'_j,t)
\dots \B.w(\B.r_n|\B.r'_n,t)\rangle.
\end{eqnarray}
Using the Navier-Stokes equations for the time derivatives we derive a
new balance equation $\B.{\cal D}^{(2)}_n+\B.{\cal B}^{(2)}_n=\B.{\cal
  J}^{(2)}_n$ where, using the definition (\ref{Lb}),
\begin{eqnarray}
&&\B.{\cal D}_n^{(2)}=\int d\B.r d\B.r' \sum_{i,j=1}^n 
 \B.P(\B.r)\B.P(\B.r')
 \Big\langle \B.w(\B.r_1|\B.r'_1)\label{D2n}\\
&&\dots \B.L(\B.r_i,\B.r'_i,\B.r)\dots \B.L(\B.r_j,\B.r'_j,\B.r')
\dots \B.w(\B.r_n|\B.r'_n)\Big \rangle \ .\nonumber
\end{eqnarray}
Using the fusion rules and following steps similar to those described
above, we can prove that the integrals over $\B.r$ and $\B.r'$
converge. Accordingly, when all the separations are of the order of
$R$, every term in ${\cal D}^{(2)}_n$ is evaluated as
$S_{n+2}(R)/R^2$.  The term $\B.{\cal J}^{(2)}_n$ takes on the form
\begin{eqnarray}
&&\B.{\cal J}^{(2)}_n=\nu^2\sum_{i,j=1}^n
\left(\nabla_i^2+\nabla_{i'}^2\right)
\left(\nabla_j^2+\nabla_{j'}^2\right)\label{J2n}\\
&\times&\langle\B.w(\B.r_1|\B.r'_1)\dots \B.w(\B.r_i|\B.r'_i)\dots
\B.w(\B.r_j|\B.r'_j)
\dots \B.w(\B.r_n|\B.r'_n)\rangle \ .\nonumber
\end{eqnarray}
As before, when all the separation in this quantity are of the order
of $R$, the Laplacian operators introduce factor of $1/R^2$ and the
evaluation of this quantity is $\B.{\cal J}^{(2)}_n\sim
\nu^2S_n(R)/R^4$. Clearly this is negligible compared to typical terms
in ${\cal D}^{(2)}_n$. The quantity $\B.{\cal B}^{(2)}_n$ contains a
cross contribution with one Laplacian operator and one nonlinear term
with a projection operator. The integral is again local, and one can
show that the evaluation is $\B.{\cal B}^{(2)}_n\sim \nu
S_{n+1}(R)/R^3$ which is also negligible compared to typical terms in
${\cal D}^{(2)}_n$.

Now we consider the fusion of two pairs of coordinate, e.g. $r_{12}\to
0$ and $r_{34}\to 0$. As before, the cancellations in $\B.{\cal
  D}^{(2)}_n$ are eliminated, and the evaluation of a typical term
becomes the evaluation of the quantity. The other two terms in the
balance equation also become of the same order because the Laplacian
operators $\nabla_1^2$ and $\nabla_3^2$ are evaluated as $r_{12}^{-2}$
and $r_{34}^{-2}$ respectively. As before we can consider the
resulting balance equation as a differential equation for
$F_n(r_{12},r_{34},\{R\})$. The leading term in this equation is
$$
4\nu^2\nabla_1^2  \nabla_2^2 F_n(r_{12},r_{34},\{R\})
\approx \B.{\cal B}^{(2)}_n
+\B.{\cal D}^{(2)}_n \sim S_{n+2}(R)/R^2 .
$$
The solution is
\begin{equation}
{\cal F}_n(r_{12},r_{34},\{R\})\sim r_{12}^2 
 r_{34}^2S_{n+2}(R)/ \nu^2 R^2 \ .
\label{dissF2n}
\end{equation}
Finally we can write the quantities $\B.{\cal
  K}_{\epsilon\epsilon}^{(n)}$ in terms of the correlation function as
\begin{equation}
\B.{\cal K}_{2\epsilon}^{(n)}=\nu^2\lim_{r_{12},r_{34}\to 0}\B.\nabla_1
\B.\nabla_2\B.\nabla_3\B.\nabla_4
\B.{\cal F}_{n+4}(r_{12},r_{34},\{R\}).\label{KF1}
\end{equation}
Using (\ref{dissF2n}) here we end up with the evaluation
\begin{equation}
\B.{\cal K}_{2\epsilon}^{(n)}
\sim S_{n+6}/R^2\propto R^{-\mu^{(2)}_n},\quad
\mu^{(2)}_n=2-\zeta_{n+6} . \label{mun}
\end{equation}
For the standard exponent $\mu=\mu_0^{(2)}$ we choose $n=0$ and obtain
the phenomenologically proposed ``bridge relation"
\begin{equation}
\mu=2-\zeta_6 \ . \label{bridge2}
\end{equation}
To our best knowledge this is the first solid derivation of this
scaling relation. In general, if we have $p$ dissipation fields
correlated with $n$ velocity differences the scaling exponent can be
found by considering $p$ time derivatives of (\ref{defF}), with the
final result
\begin{equation}
\mu^{(p)}_n=p-\zeta_{n+3p}.\label{mup}
\end{equation}
We see that Eqs.(\ref{mu1n}), (\ref{mun}) and (\ref{mup}) can be
guessed if we assert that {\em for the sake of scaling purposes} the
dissipation field $\epsilon(\B.r)$ can be swapped in the correlation
function with $w^3(\B.r_1|\B.r'_1)/R_1$, where $R_1$ is the
characteristic scale. This reminds one of the Kolmogorov refined
similarity {\em hypothesis}. We should stress that (i) our result does
not depend on any uncontrolled hypothesis, and (ii) it does not imply
the correctness of the hypothesis. Our result is implied by the
refined similarity hypothesis, but not vice versa.

We end this section by noting that the accepted values of $\mu$ and
$\zeta_6$ which are about $0.2$ and $1.8$ respectively, are in good
agreement with the standard bridge relation (\ref{bridge2}). However,
we have presented here a "two-dimensional" array of bridge relations
depending on the indices $n$ and $p$ whose experimental test with high
Reynolds number flows with good resolution of the dissipative scales
is highly desirable, considering the putative exact nature of these
relations.
\section{Summary and Conclusions}
In terms of new predictions the theory described above has a lot to
offer.  Firstly, we have presented the fusion rules, and it is
extremely worthwhile to test them against experimental data. One can
achieve a reasonable test already by using existing data sets from
atmospheric boundary layers or grid turbulence. In such experiments
one measures usually at one space point as a function of time. Using
the standard Taylor hypothesis one can measure many-point correlation
functions for points placed along one line. It is possible to examine
the properties of such correlation functions when one distance is much
smaller than all others.  Another prediction pertains to the viscous
scaling function and the anomalous exponents $x_n$ that characterize
them, see Eq.(\ref{dissFn}. To test these predictions one needs a good
resolution of the sub-dissipative scales in a high Reynolds number
experiment. Such data are not readily available, but very worthwhile
to acquire. Another important point raised briefly in this paper has
to do with the set of exponents $\beta_l$ which govern the anisotropic
properties of the correlation functions. We noted elsewhere that the
same set of exponents characterizes the correlation functions
\acknowledgments This work was supported in part by the German Israeli
Foundation, the US-Israel Bi-National Science Foundation, the Minerva
Center for Nonlinear Physics, and the Naftali and Anna
Backenroth-Bronicki Fund for Research in Chaos and Complexity.
\appendix
\section{Some special
 geometries of fusion and their implications}
\label{sub-D}
\subsection{Special geometry of Fig.~\protect\ref{Fig4}}
\label{fuse-Fig4}
\begin{figure}
\label{Fig4}
\end{figure}

{\small FIG.~\ref{Fig4}. Special geometry of fusion that is discussed
  in Appendix A 1}\vskip .4cm

As we discussed in subsection \ref{fusep1}, the evaluation
(\ref{rule2}) is inapplicable when the smallest of the large
separations (say $|\B.r_1-\B.r_2|=R_{\rm min}$) (which is not
associated with a velocity difference) becomes similar to small
separations $r$ (across a velocity difference). To understand how to
evaluate the correlation function in this case consider the limit
$R_{\rm min}=0$ with the help of the special geometry shown in
Fig.~\ref{Fig4}.  We have four coordinates, $\B.r_1$, $\B.r'_1$,
$\B.r_2$ and $\B.r'_2$ organized as shown in the figure, i.e.
$\B.r'_1-\B.r_1$ along the $z$-axis, $\B.r_2=\B.r_1$, and $\B.r'_2$ is
on the $x$-axis, with equal distances to $\B.r_1$ and $\B.r'_1$. This
special geometry will help us to derive a result that holds more
generally. In addition to the velocity differences
$\B.w(\B.r_1,\B.r'_1)$ and $\B.w(\B.r_2,\B.r'_2)$ we can have any
number of of velocity differences $\B.w(\B.r_i,\B.r'_i)$, but we
demand that the product of all the additional $n-2$ velocity
differences (denoted in short-hand as $\{w\}^{n-2}$) remains invariant
to rotations of $\pi$ radians around the $x$ axis. In Fig.~\ref{Fig4}
we show two such additional velocity differences across
$\B.r_3-\B.r'_3$ and $\B.r_4-\B.r'_4$.  We will write the $n$-th order
correlation function in short-hand notation as
\begin{eqnarray}
&&F_n({\B.r}_1,{\B.r}'_1;{\B.r}_2,{\B.r}'_2; \{{\B.r}_k,{\B.r}'_k\})
\nonumber \\
&=&
\left<\B.w(\B.r_1,\B.r'_1)\cdot\B.w(\B.r_2,\B.r'_2)\{w\}^{n-2}\right>
\ . \label{defsymF}
\end{eqnarray}
By definition $\B.w(\B.r_2,\B.r'_2)=
\B.w(\B.r_2,\B.r'_1)+\B.w(\B.r'_1,\B.r'_2)$.  Remembering that
$\B.r_1=\B.r_2$ we get
\begin{eqnarray}
&&F_n({\B.r}_1,{\B.r}'_1;{\B.r}_2,{\B.r}'_2; \{{\B.r}_k,{\B.r}'_k\})=
F_n({\B.r}_1,{\B.r}'_1;{\B.r}'_1,{\B.r}'_2; \{{\B.r}_k,{\bf
r}'_k\})\nonumber \\
&&+
F_n({\B.r}_1,{\B.r}'_1;{\B.r}_1,{\B.r}'_1; \{{\B.r}_k,{\B.r}'_k\})\ .
\label{FFF}
\end{eqnarray}
Due to the rotation symmetry the first term on the right hand side
(RHS) of (\ref{FFF}) may be written as $F_n({\B.r}_1,{\B.r}'_1;{\bf
  r}_1,{\B.r}'_2; \{{\B.r}_k,{\B.r}'_k\})$, which by definition is the
$-F_n({\B.r}_1,{\B.r}'_1;{\B.r}_2,{\B.r}'_2; \{{\B.r}_k,{\bf
  r}'_k\})$. Finally we derive the identity
\begin{equation}
F_n({\B.r}_1,{\B.r}'_1;{\B.r}_2,{\B.r}'_2; \{{\B.r}_k,{\B.r}'_k\})=
\case{1}{2} F_n({\B.r}_1,{\B.r}'_1;{\B.r}_1,{\B.r}'_1; \{{\B.r}_k,{\bf
r}'_k\}) .
\label{ident}
\end{equation}
The object on the RHS has two velocity differences across a small
separation $r$, and $n-2$ large separations. Therefore it follows the
usual fusion rules (\ref{fusion1}) for $p=2$. Accordingly
\begin{eqnarray}
&& {\B.F}_n({\B.r}_1,{\B.r}'_1;{\B.r}_2,{\B.r}'_2; \{{\B.r}_k,{\bf
r}'_k\}) \nonumber \\
&=&
{\tilde{\B.F}}_2({\B.r}_1,{\B.r}'_1) 
\bbox\Psi_{n,2}\{{\B.r}_k,{\B.r}'_k\})
\propto \left({r\over R}\right)^{\zeta_2}R^{\zeta_n}\ . \label{funny}
\end{eqnarray}
This result was derived for the very specific geometry shown in
Fig.~\ref{Fig4}.  However, the last formula holds much more generally.
Even if we tilt the vector $\B.r'_1-\B.r_1$ in an arbitrary angle, the
result remains invariant. The reason is that in the limit $r\ll R$ the
anisotropic effects of the large scales on the small scales have
already disappeared, as is shown in Section III. If we ruin the
symmetry of rotation around the $x$-axis we also do not change the
final result.  The change will be in the factor in Eq.(\ref{ident})
from $1/2$ to a geometry-dependent factor of the order of unity. We
can even dissociate $\B.r_2$ from $\B.r_1$ over distances of the order
of $|\B.r'_1-\B.r_1|$. Eq.(\ref{funny}) is rather universal.
\subsection{Special geometry of Fig.~\protect\ref{Fig5}}
\begin{figure}
\label{Fig5}
\end{figure}

{\small FIG.~\ref{Fig5}. Special geometry of fusion that is discussed
  in Appendix A 2}\vskip .4cm

In Fig.~\ref{Fig5} we show a situation in which there exist two
velocity differences across short distances. The difference with the
general situation shown in Fig.~1 (for $p=2$) is that we have a fusion
of points belonging to short and long separations, i.e
$\B.r_2=\B.r_3$.  The same type of fusion existed also in
Fig.~\ref{Fig4}, but there was only one one short distance with a
velocity difference across it. Now we have two.  According to the
general rule we should have a contribution that is proportional to
$r^{\zeta_2}$. The aim of this special discussion is to show that in
this case there exists a subleading contribution that is proportional
to $r^{\zeta_3}$. This is important in the analysis that involves the
calculation of gradients with respect to these positions.  To this aim
consider the correlation function
\begin{eqnarray}
&&F_n({\B.r}_1,{\B.r}'_1;{\B.r}_2,{\B.r}'_2; {\B.r}_3,{\B.r}'_3;
\{ {\B.r}_k,{\B.r}'_k\})\label{defsym3F} \\
&=&
\left<\big[\B.w(\B.r_1,\B.r'_1)\cdot\B.w(\B.r_2,\B.r'_2)
\big]\big[\B.w(\B.r_3,\B.r'_3)\cdot \hat {\B.z}\big]
\{w\}^{n-3}\right>
\nonumber 
\end{eqnarray}
where $\hat {\B.z}$ is a unit vector in the $z$ direction. We again
use a short hand notation $\{w\}^{n-3}$ for the product of $n-3$
velocity differences across large separations which depend on the
coordinates $\B.r_4$, $\B.r'_4$ and higher.  Using now the fact that
$\B.w(\B.r_3,\B.r'_3)=\B.w(\B.r_3,\B.r'_1)+ \B.w(\B.r'_1,\B.r'_3)$ we
compute
\begin{eqnarray}
&&F_n({\B.r}_1,{\B.r}'_1;{\B.r}_2,{\B.r}'_2; {\B.r}_3,{\B.r}'_3;
\{{\B.r}_k,{\B.r}'_k\}) \label{defsym4F}\\
&=&
\left<\big[\B.w(\B.r_1,\B.r'_1)\cdot\B.w(\B.r_2,\B.r'_2)
\big] \big[\B.w(\B.r'_1,\B.r'_3)
\cdot \hat {\B.z}\big]\{w\}^{n-3}\right>
\nonumber\\ &&+\left<\big[\B.w(\B.r_1,\B.r'_1)
\cdot\B.w(\B.r_2,\B.r'_2)
\big] \big[\B.w(\B.r_3,\B.r'_1)\cdot \hat {\B.z}\big] 
\{w\}^{n-3}\right>
\ .\nonumber 
\end{eqnarray}
Rotating around the $x$-axis we can rewrite the last correlator on the
RHS of (\ref{defsym4F}) as the correlator on the LHS with an opposite
sign.  To see this note that all the terms are invariant except the
term $\B.w(\B.r_3,\B.r'_1)\cdot \hat {\bf z}$ that changes sign. Thus
finally
\begin{eqnarray}
&&F_n({\B.r}_1,{\B.r}'_1;{\B.r}_2,{\B.r}'_2; {\B.r}_3,{\B.r}'_3;
\{{\B.r}_k,{\B.r}'_k\})\\
&=&\case{1}{2}
\left<\big[\B.w(\B.r_1,\B.r'_1)\cdot\B.w(\B.r_2,\B.r'_2)
\big] \big[ \B.w(\B.r'_1,\B.r'_3)\cdot \hat {\B.z}\big] 
\{w\}^{n-3}\right>\nonumber 
\end{eqnarray}
This correlator has three explicit velocity differences across short
distances, and therefore according to the general rule with $p=3$ it
is proportional to $r^{\zeta_3}$. In the general case without rotation
symmetry the leading term $r^{\zeta_2}$ remains. Therefore we conclude
that in the geometry in which there are two velocity differences
across a small separation and one point belonging to a velocity
difference across a large separation (see Fig.~\ref{Fig5}), the
correlation function $F_n$ can be written to leading order as
\begin{eqnarray} 
&&\B.F_n({\B.r}_1,{\B.r}'_1;{\B.r}_2,{\B.r}'_2; {\B.r}_3,{\B.r}'_3;
\{{\B.r}_k,{\B.r}'_k\})=[\tilde\B.S_2(|\B.r_1\!-\!\B.r_2|)
\label{hori} \\&&+\tilde\B.S_2(|\B.r'_1\!-\!\B.r'_2|)
 -\tilde\B.S_2(|\B.r_1\!-\!\B.r'_2|)
-\tilde\B.S_2(|\B.r'_1\!-\!\B.r_2|)]\nonumber\\
&&\B.\Psi_{n,2}(\B.r_0,\B.r'_3;
\{{\B.r}_k,{\B.r}'_k\})+[\tilde \B.S_3(\B.r_0|\B.r_1,\B.r_2,\B.r_3)
+\tilde \B.S_3(\B.r_0|\B.r'_1,\B.r'_2,\B.r_3)\nonumber \\&&-
\tilde \B.S_3(\B.r_0|\B.r'_1,\B.r_2,\B.r_3)
-\tilde \B.S_3(\B.r_0|\B.r_1,\B.r'_2,\B.r_3)]
\B.\Psi_{n,3}(\B.r_0,\{{\B.r}_k,{\B.r}'_k\}) \ ,\nonumber 
\end{eqnarray}
\noindent
where $\B.r_0=[\B.r_1+\B.r'_1+\B.r_2+\B.r'_2+\B.r_3]/5$, and $\tilde
\B.S_2$, $\tilde\B.S_3$, $\B.\Psi_{n,2}$ and $\B.\Psi_{n,3}$ are
homogeneous functions of their arguments (in the inertial interval)
with scaling exponents $\zeta_2$, $\zeta_3$, $\zeta_n-\zeta_2$ and
$\zeta_n-\zeta_3$ respectively. The function $\tilde \B.S_3$ may be
different from the function $S_3$ in its dependence on the angles and
the ratios between its argument coordinates. But they share the same
scaling exponent.
\subsection{Special geometry of Fig.~\protect\ref{Fig7}}
\begin{figure}
\label{Fig7}
\end{figure}

{\small FIG~\ref{Fig7}. Special geometry of fusion that is discussed
  in Appendix A 3}\vskip .4cm

In this subsection we consider the case in which there are four
coordinates within the ball of size $r$, but only two coordinates
($\B.r_1$ and $\B.r'_1$) belong to a velocity difference, see
Fig.~\ref{Fig6}.  The other two coordinates ($\B.r_2$ and $\B.r_3$)
are in the ball, but they relate to velocity differences across large
separations. To understand the situation we again consider a special
geometry, that of Fig.~\ref{Fig7}. In this geometry
$\B.r_1=\B.r_2=\B.r_3$ and we study the correlation function
\begin{eqnarray}
&&\B.F_n({\B.r}_1,{\B.r}'_1;{\B.r}_2,{\B.r}'_2; {\B.r}_3,{\B.r}'_3;
\{{\B.r}_k,{\B.r}'_k\})
=\langle\big[\B.w(\B.r_1,\B.r'_1)\cdot\hat {\B.z}\big]
\nonumber \\ &&\big[\B.w(\B.r_2,\B.r'_2)
\cdot\B.w(\B.r_3,\B.r'_3)\big]\{w\}^{n-3}\rangle
\ . \label{odF} 
\end{eqnarray}
Making the substitutions
\begin{eqnarray}
\B.w(\B.r_2,\B.r'_2)=\B.w(\B.r_1,\B.r'_2)
&=&\B.w(\B.r_1,\B.r'_1)+\B.w(\B.r'_1,\B.r'_2)\\
\B.w(\B.r_3,\B.r'_3)=\B.w(\B.r_1,\B.r'_3)
&=&\B.w(\B.r_1,\B.r'_1)+\B.w(\B.r'_1,\B.r'_3)
\end{eqnarray}
we find that
\begin{eqnarray}
&2&\B.F_n({\B.r}_1,{\B.r}'_1;{\B.r}_2,{\B.r}'_2; {\B.r}_3,{\B.r}'_3;
\{ {\B.r}_k,{\B.r}'_k \} )=\langle \big[\B.w(\B.r_1,\B.r'_1)
\cdot\hat {\B.z}\big]\nonumber\\
&\times&\big|\B.w(\B.r_1,\B.r'_1)\big|^2 
\{w\}^{n-3}\rangle +\langle [\B.w(\B.r_1,\B.r'_1)\cdot\hat 
{\B.z}]\nonumber\\
&\times&\B.w(\B.r_1,\B.r'_1) 
\cdot[\B.w(\B.r'_1,\B.r'_2)+\B.w(\B.r'_1,\B.r'_3)]\{w\}^{n-3}\rangle
\ . \label{odF1}
\end{eqnarray}
In obtaining this equation we used the symmetry under rotation around
the $x$ axis in $\pi$. Under this rotation $\B.r_1\to\B.r'_1$. The
first term on the RHS has an explicit product of three velocity
differences across a small distance, and it is therefore proportional
to $r^{\zeta_3}$. The second term has two velocity differences across
a small scale, and according to the previous subsection it contains
two contributions, one proportional to $r^{\zeta_2}$ and the other to
$r^{\zeta_3}$.  One can write it in a form similar to (39) and (40),
but this is not needed at the moment.

\end{multicols}
\end{document}